\documentclass[reprint,amsfonts, amssymb, amsmath, showkeys,aps,pra, superscriptaddress, twocolumn,longbibliography,nofootinbib]{revtex4-2}

\newcommand{\doctitle}{Distributions of noisy expectation values over sets of measurement operators}
\newcommand{\docauthor}{Matthew Duschenes}
\newcommand{\docauthorone}{Roger G. Melko}
\newcommand{\docauthortwo}{Juan Carrasquilla}
\newcommand{\docauthorthree}{Raymond Laflamme}
\newcommand{\docaffil}{Department of Physics \& Astronomy, University of Waterloo, Ontario, N2L 3G1, Canada}
\newcommand{\docaffilone}{Institute for Quantum Computing, University of Waterloo, Ontario, N2L 3G1, Canada}

\newcommand{\docaffilthree}{Perimeter Institute for Theoretical Physics, Waterloo, Ontario, N2L 2Y5, Canada}
\newcommand{\docaffilfour}{Institute for Theoretical Physics, ETH Zürich, 8093, Switzerland}
\newcommand{\docemail}{mduschen@uwaterloo.ca}
\newcommand{\docemailone}{Deceased}

\newcommand{\pdftitle}{\doctitle}

\usepackage{main}

\begin{document}

\preprint{APS/123-QED}

\title{\doctitle}

\author{\docauthor}
\email[Corresponding author: ]{\docemail}
\affiliation{\docaffil}
\affiliation{\docaffilone}
\affiliation{\docaffilthree}
\author{\docauthorone}
\affiliation{\docaffil}
\affiliation{\docaffilthree}
\author{\docauthortwo}
\affiliation{\docaffilfour}
\affiliation{\docaffil}
\author{\docauthorthree}
\email{\docemailone}
\affiliation{\docaffil}
\affiliation{\docaffilone}
\affiliation{\docaffilthree}

\date{\today}

\begin{abstract}
Expectation values of measurement operators, interpreted as measurement probabilities, arise frequently throughout quantum information sciences. When quantum states are randomly distributed, their expectation values are also randomly distributed. In this work, with the goal of understanding non-unitary dynamics, we generalize previous derivations for distributions of expectation values [Campos \emph{et al}., Phys. Lett. A 377, 1854 (2013)] to the case of sets of measurement operators and random mixed quantum states within variable sized environments. Using combinatorics approaches, we derive expressions for their moments. We proceed to construct empirical distributions of simulated Haar random brickwork quantum circuits with local depolarizing noise, and compare their form to a proposed effective global-depolarizing-like model with variable effective noise scales and environment dimensions. The fitted effective distributions reproduce peak behavior across circuit depths, noise scales, and system sizes, while deviations in the distribution tails arise from local noise effects. The fit effective model parameters are also shown to vary smoothly and consistently with circuit depth and noise scale. Finally, sets of nonsymmetric measurement operators are shown to exhibit distinct multimodal distributions relative to unimodal distributions for symmetric measurement operators, opening up questions about their simulability.
\end{abstract}

\maketitle

\section{Introduction}\label{sec:introduction}

Expectation values of systems described by quantum states, given quantum theory is inherently probabilistic in nature \cite{appleby2009properties,appleby2016introducing}, are at the heart of quantum information sciences. Experimental probes, numerical simulations, or analytical studies thus involve measurement procedures that allow us to interact with and extract information from a system of interest. Expectation values can have a variety of interpretations. Expectation values of measurement operators specify the probabilities of measurement outcomes \cite{scott2006tight,yashin2020minimal,renes2003symmetric}, expectation values of Hamiltonians determine average system energies \cite{miller2025statistical,yashin2020minimal}, and fidelities quantify distances between quantum states \cite{zyczkowski2003average,sommers2005bures,meyer2021fisher,zhou2020what}.

Randomness in quantum states \cite{sommers2004statistical,sommers2005bures} arises due to intrinsic stochastic dynamics \cite{fefferman2024effect,holmes2021connecting,duschenes2025moments},  algorithmic design \cite{emerson2005scalable,quek2022exponentially,wang2021can,emerson2007symmetrized}, and in the description of chaos and thermalization in quantum many-body systems \cite{srednicki1994chaos}. Randomness is also a key concept in random circuit sampling experiments to compare classical and quantum resources \cite{fefferman2024effect,zhang2023noisy,fisher2023random,cheng2023efficient,czischek2021simulating,arute2019quantum,boixo2018characterizing,larose2024brief,kim2023evidence,zhu2022quantum}, tomography to efficiently learn system properties \cite{torlai2023quantum,huang2020predicting,carrasquilla2019reconstructing,hu2021classical,hu2023tackling}, concentration phenomena within variational algorithms \cite{holmes2021connecting,duschenes2025moments,larocca2024review}, optimal control \cite{duschenes2024characterization,sivak2023real,georgopoulos2021modelling,ge2022optimization}, and error mitigation \cite{tsubouchi2025symmetric,jnane2024quantum,wang2021can,bulchandani2024random,quek2022exponentially,emerson2007symmetrized}. Random quantum circuits consisting of layers of unitaries followed by measurements, give rise to measurement induced phase transitions \cite{fisher2023random}. As the probability of local measurements is increased, states can transition from being volume-law entangled, to being area-law entangled, with well-defined critical exponents and universality classes \cite{skinner2019measurement,li2018quantum}.

More concretely, given states that are randomly distributed from a known distribution, their properties will themselves also be randomly distributed, from a yet-to-be-determined distribution. Expectation values are also always calculated with respect to specific states and operators, and of particular relevance to quantum computing are measurement operators \cite{heinosaari2019random,appleby2016introducing,appleby2009properties}. It is thus essential to understand these expectation value distributions, given specific operators, and ensembles of states.

When sets of measurement operators are so-called informationally complete \cite{scott2006tight,renes2003symmetric,acharya2021informationally}, meaning their measurement probabilities for distinct quantum states are uniquely defined, then consistent tomography can be performed \cite{anshu2023survey,thew2002qudit,gross2010quantum,cramer2010efficient,carrasquilla2019reconstructing,torlai2023quantum}. General informationally complete measurement procedures have far-reaching applications, such as classical shadow tomography techniques \cite{huang2020predicting,hu2021classical,koh2020classical,acharya2021informationally} and measurement-induced phase transitions \cite{skinner2019measurement,li2018quantum}. By posing such applications in terms of full distributions rather than expectation values alone, it is possible to obtain a more complete characterization of how observables depend on system parameters such as system size or evolution time. Further, when considering distributions over sets of operators \cite{heinosaari2019random}, total distributions of expectation values noticeably become sums over conditional distributions, conditioned on the choice of operator. Such total distributions, and their dependence on any defining properties of these sets of operators, have yet to be systematically explored.

Significant progress has been made in understanding the distributions of expectation values of projective measurements for pure, uniformly distributed states. Such settings give rise to the famous Porter-Thomas distribution \cite{mullane2020sampling,claeys2024fock}, and mixed state variants of this distribution are explored throughout this work. Here, notions of uniformity imply dynamics are distributed according to the unitarily-invariant Haar measure \cite{collins2022weingarten,kukulski2021generating}. Crucially, a quantum circuit's architecture that dictates these dynamics, and any global versus local effects, greatly affect ease of analysis \cite{belkin2024approximate,deshpande2022tight,kus1988universality,dalzell2022randomquantum}.

In fact, closed-form expressions have been derived for distributions of expectation values of Hermitian operators with respect to Haar random pure states \cite{camposvenuti2013probability,dunkl2011numerical,gutkin2013joint,zhang2021uncertainty}. Recently, more experimentally realistic \cite{kim2023evidence,sivak2023real,czischek2021simulating} locally random quantum circuits with brickwork layouts \cite{sauliere2025chaotic,brown2010random,belkin2024approximate}, and projective measurements, have been investigated. In such analyses, it is typically intractable to compute closed-form expressions for distributions, and instead moments or other statistics are computed \cite{sauliere2025chaotic,sauliere2026universality,loio2025correlations,deluca2025universality} using Weingarten calculus \cite{collins2022weingarten,collins2017weingarten,collins2006integration}. Subsequently, distributions, approximated using these moments, are shown to be in good agreement with empirical histograms. However, without closed form expressions, quantitatively interpreting such distributions as a function of various system parameters remains difficult.

Further complicating these studies is when systems undergo noisy dynamics \cite{duschenes2024characterization,bharti2022noisy,cheng2023efficient,zhang2023noisy,koh2020classical,lee2024universal,fefferman2024effect}, or directly interact in open settings with their environment \cite{duschenes2025moments,sang2023mixed,zhang2022entanglement}. The subtle interplay between the effects of noise with specific noise scales \cite{cheng2023efficient,denzler2026simulation} or environment dimensions \cite{duschenes2025moments,bulchandani2024random,bai2024primitivity,kukulski2021generating}, versus the effects of entangling unitary dynamics \cite{fisher2023random}, is just beginning to be studied, using insights from random matrix theory \cite{sauliere2026universality,collins2016random,zyczkowski2001induced} and variational quantum algorithms \cite{holmes2021connecting,duschenes2025moments,collins2009random}.

In this work, we seek to contribute to answering the following: How do parameters such as system size, circuit depth, and noise scale affect distributions of expectation values of sets of general measurement operators, and given insight into distributions for globally noisy random systems, are there effective analytical models that capture behaviors of locally noisy random systems?

In \cref{sec:preliminaries}, we introduce our formalisms to describe states, operators, and their statistics. In \cref{sec:results}, we extend the previously derived distributions of expectation values to the case of Haar random mixed states with nontrivial environment dimensions, and by using combinatorics approaches, we derive expressions for their moments. We proceed to form empirical distributions of symmetric and nonsymmetric positive-operator-valued (POVM) measurement probabilities from simulated noisy brickwork circuits, and propose an effective analytical model with interpretable variable parameters. By using metrics to distinguish continuous variable distributions, we demonstrate the suitability of these effective models at describing peak versus tails behaviors. We find that these distributions develop sharp peaks with increased circuit depth and noise scale, become unimodal for symmetric measurements, and multimodal for nonsymmetric measurements. Finally in \cref{sec:discussion}, we show how the fit effective model parameters depend on system parameters of system size, circuit depth, and noise scale.

\section{Preliminaries}\label{sec:preliminaries}

Here, we introduce formalisms used to study expectation values of Hermitian operators, with respect to quantum states within $d$-dimensional spaces, and $s$-dimensional environments. After introducing preliminary notation, we will introduce distributions of various quantities, before discussing metrics for numerically quantifying differences between such distributions.

\subsection{Quantum states, dynamics, operators, and expectation values}
We consider $s$-rank $d$-dimensional quantum states $\rho$, which can be parameterized by $d \times s$-dimensional parameters $\varphi$,
\begin{align}
	\rho = \varphi\varphi^{\dagger}~,
\end{align}
as well as trace-preserving completely positive maps, or quantum channels,
\begin{align}
	\Lambda = \prod_{i \in [k]} \Lambda^{(i|k)}
	\quad : \quad
	\rho \to \Lambda(\rho)
	~,
\end{align}
with  $k$ layers indexed by $i \in [k] = \{0,1,\dots,k-1\}$.

Regarding operators, we consider Hermitian $l$-rank operators, with $\#$ distinct real eigenvalues $\{\sigma \leq \xi \leq \lambda\}$,
\begin{align}
	\Pi = \sum_{\xi} \xi~I_{\xi}
	\quad : \quad \sum_{\xi}I_{\xi} = I
	~,
\end{align}
with $d_{\xi}$-dimensional eigenspaces, with projectors $I_{\xi}$.

Given such states, dynamics, and operators, here we consider expectation values $x$ of operators $\Pi$, with respect to states $\rho$, defined via the linear maps,
\begin{align}
	\rho \to \tau_{\Pi}(\rho) = \trace{\Pi~\rho} ~\equiv~ x~.
\end{align}

Operators and their expectation values are bounded by their maximum $\lambda$ and minimum $\sigma$ eigenvalues,
\begin{align}
	\sigma I \leq \Pi \leq \lambda I
	\quad \quad , \quad \quad
	\sigma \leq x \leq \lambda~.
\end{align}
We find it convenient to normalize operators,
\begin{align}
	\Pi \to \frac{\Pi - \sigma I}{\lambda - \sigma}
	\quad , \quad
	x \to \frac{x-\sigma}{\lambda-\sigma}~,
\end{align}
which are bounded by their eigenvalues,
\begin{align}
	0 \leq \Pi \leq I
	\quad \quad , \quad \quad
	0 \leq x \leq 1~.
\end{align}

$\Pi \equiv \Gamma^{\dagger}\Gamma \geq 0$ are positive, rank $0\leq l\leq d$, with $\#$ number of distinct non-negative eigenvalues. The image of $\Pi$ induces $l,d-l$-dimensional subspaces with associated $l,d-l$-rank projectors $I_{\Pi},I-I_{\Pi}$.

States $\rho$, operators $\Pi$, and expectation values $x$ can be jointly transformed by hermiticity-preserving trace-preserving maps, such as quantum channels $\Lambda$ and their associated adjoints $\Lambda^{\dagger}$, with the mappings,
\begin{align}
	\rho \to \rho_{\Lambda} = \Lambda(\rho)
	\quad\leftrightarrow&\quad
	\Pi \to \Pi_{\Lambda^{\dagger}} = \Lambda^{\dagger}(\Pi) \\
	x \to x_{\Lambda} = \tau_{\Pi}&~\!(\rho_{\Lambda}) = \tau_{\Pi_{\Lambda^{\dagger}}}(\rho) ~.
\end{align}

\subsection{Distributions of expectation values of operators}
Here, let us build up intuition for our choice of distributions of states and expectation values, and let us derive a useful invariance property, as detailed in \cref{sec:distributions_of_expectation_values_of_operators}.

More concretely, given fixed operators $\Pi$, and randomly distributed states $\rho \sim P_{\rho}$ or parameters $\varphi \sim P_{\varphi}$,
\begin{align}
	P_{\rho}(\rho)
	=&~ \int d\varphi~P_{\varphi}(\varphi)~\dirac{\rho-\varphi\varphi^{\dagger}} ~,
\end{align}
expectation values $x \sim P_{\Pi}$ are thus also randomly distributed, with distributions,
\begin{align}
	P_{\Pi}(x)
	=&~ \int d\rho~P_{\rho}(\rho)~\dirac{x-\tau_{\Pi}(\rho)}~.
\end{align}

Let us now make some assumptions about the distribution of states and parameters, with the objective of studying approximately uniformly random states \cite{zyczkowski2001induced,zyczkowski2011generating,collins2016random}. Unlike in the pure state case, the general mixed state case has additional freedom in choosing distributions for both the state eigenvectors and the eigenvalues independently. $s$-rank mixed states are thus only uniformly distributed when $s=d$ \cite{zyczkowski2001induced}. Here we will assume unitary invariance, such that our distribution of mixed states is induced by a uniform distribution of pure states in the composite system and environment space.

Given the unitary invariance of the measure $d\varphi$, we will assume the distribution $P_{\varphi} = P_{\varphi}(\norm{\varphi})$ is also unitarily invariant, and is isomorphic to the Haar measure of $ds$-dimensional pure states $d\psi$, with solely the norm constraint, given $2ds$-dimensional areas $\Omega_{2ds}$,
\begin{align}
	P_{\varphi}(\varphi)~d\varphi ~~=~~ \frac{1}{\frac{1}{2}\Omega_{2ds}} \dirac{\norm{\varphi}^{2}-1}d\varphi ~~\cong~~ d\psi~.
\end{align}

Given such unitary invariance and constraints, we find that expectation value distributions have shift and scale invariance, and thus unnormalized and normalized operator distributions are related by,
\begin{align}
	P_{\Pi}(x) = \displaystyle\frac{1}{\lambda-\sigma}P_{\frac{\Pi - \sigma I}{\lambda-\sigma}}\left(\frac{x-\sigma}{\lambda-\sigma}\right) . \!
\end{align}

For example, suppose we have a global depolarizing channel with noise scale $\gamma$, then states, operators, and distributions are shifted and scaled as,
\begin{align}
	\rho \to \rho_{\gamma} =(1-\gamma)\rho ~+&~  \gamma\frac{\trace{\rho}}{d}I \\
	\leftrightarrow&~ \nonumber \\
	\Pi \to \Pi_{\gamma} = (1-\gamma)\Pi ~+&~  \gamma\frac{\trace{\Pi}}{d}I \\
	x \to x_{\gamma} = (1-\gamma)x ~+&~ \gamma\frac{\trace{\Pi}}{d} \\
	P_{\Pi_{\gamma}}(x_{\gamma}) = \displaystyle\frac{1}{1-\gamma}&~P_{\Pi}\left(\frac{x_{\gamma} - \gamma\trace{\Pi}/d}{1-\gamma}\right) . \!
\end{align}

We seek to understand such distributions of expectation values, namely their analytical form, and their similarity to empirical distributions from simulated systems.

\subsection{Metrics quantifying differences in distributions}
Given an analytical distribution $P$ or cumulative distribution $F$, $m$ independent and identically distributed samples $\{x_{i} \sim P\}_{i \in [m]}$ form an empirical distribution,
\begin{align}
	\!\!\!\!
	\tilde{F}(x) = \frac{1}{m}\sum_{i \in [m]}\indicator{x \geq x_{i}} ~\approx~ F(x) = \int^{x}dz~P(z) ~, \!\!
\end{align}
in terms of indicator functions, $\indicator{x} = \left\{\begin{array}{cc} 1 & x~\textrm{is True}\\0 & x~\textrm{is False}\end{array}\right.$, with means and variances in terms of the distribution $F$,
\begin{align}
	\mu_{\tilde{F}}(x) = F(x)
	\quad,\quad
	\Sigma_{\tilde{F}}(x) = \frac{1}{m}F(x)~[1-F(x)]~.
\end{align}

As detailed in \cref{sec:simulations_of_expectation_values_of_operators}, a metric $\mathcal{L} = \mathcal{L}(\tilde{F},F)$ must be chosen to determine how the empirical distribution converges with system parameters to a known reference distribution $F$. However, the empirical distribution may be biased and converge to a different distribution, $\tilde{F} \to F^{\prime} \neq F$, such as in noisy, finite system size, or finite depth simulation settings.

Potential metrics vary in their relevance to discrete versus continuous variable distributions, their use of empirical versus analytic expressions, their computational complexity, and ultimately their sample complexity, such that the metrics accurately reflect differences in the distributions \cite{bobkov2010concentration}. The Kullback–Leibler divergence, being a statistical distance, is an attractive choice of metric \cite{carrasquilla2019reconstructing} in particular in discrete variable settings. However, such a metric cannot be efficiently estimated from samples in continuous variable settings \cite{perezcruz2008kullback}, and we therefore choose a statistical test that is suited for empirical distributions.

Due to its intuitive form, in this work we use the Kolmogorov–Smirnov \cite{dimitrova2020computing,bobkov2010concentration} metric, of the maximum difference between cumulative distributions,
\begin{align}
	\mathcal{L} =&~ \max_{x}~\abs{\tilde{F}(x)-F(x)}~.
\end{align}
Given the monotonicity of cumulative distributions, such a maximization over the entire domain $x$ can be made numerically tractable by being upper-bounded by an empirical metric $\mathcal{L} \leq \tilde{\mathcal{L}}$ over the $m$ samples as,
\begin{align}
	\tilde{\mathcal{L}} =&~ \max_{i \in [m]} ~\abs{\tilde{F}(x_{i})-F(x_{i})}+\abs{F(x_{i+1})-F(x_{i})}~.
\end{align}
Finally, if we assume $\tilde{F} \to F$ is unbiased, and converges with $m \to \infty$ samples, given $F=1/2$ maximizes the sample variance, we can bound the metric sample complexity using Chebyshev's inequality for the probability $\delta^{2}>0$ of quantities differing from their mean by $\epsilon>0$,
\begin{align}
	\!\!
	P\left(\abs{\tilde{F}-\mu_{\tilde{F}}} \geq \epsilon\right) \leq&~ \frac{\Sigma_{\tilde{F}}}{\epsilon^{2}} \leq \delta^{2} ~~\to~~
	m \geq \left(\frac{1}{2\delta\epsilon}\right)^{2}. \!\!
\end{align}

Given such definitions and assumptions, our goal is to derive expressions for the distribution of expectation values, and to simulate particular random quantum systems. Given analytical and empirical distributions, we can use our metrics to assess whether the distributions converge as a function of various system parameters.

\section{Results}\label{sec:results}
Here, we discuss properties of distributions of expectation values for general Hermitian operators. Such expressions are derived in \cref{sec:distributions_of_expectation_values_of_operators}, where we extend previous derivations for pure states \cite{camposvenuti2013probability,dunkl2011numerical,bulchandani2024random}, to mixed states. We then perform numerical studies of empirical distributions of measurement probabilities, given simulated noisy quantum circuits and measurement operators.

\subsection{Projection operators}
First, let us study properties for projectors,
\begin{align}
	\Pi = I_{\Pi}~,
\end{align}
with $0 \leq l \leq d$-rank, and expectation values $0 \leq x \leq 1$, that follow the elegant $l,s$-dependent Beta distributions,
\begin{align}
	P_{\Pi}(x)
	=&~ \frac{\Gamma_{ds}}{\Gamma_{ls}~\Gamma_{(d-l)s}}~x^{ls-1}~\left(1 - x\right)^{(d-l)s-1} ~,
\end{align}
with the following properties,
\begin{align}
	\!\!\!\!\textrm{Form:}& \!\!\!\!& ~&~ P_{\Pi}(x)~~
	\!\!&\to&\displaystyle\left\{\begin{array}{ll}
	x^{ls}e^{-dsx} & \substack{d \to \infty\\s \to \infty}\\[6pt]
	e^{-dx} & \substack{d \to 1\\s \to 1}
	\end{array}\right. \nonumber \!\!\\
	\!\!\!\!\textrm{Moments:}& \!\!\!\!&x_{t} =&~ \frac{\Gamma_{ds}}{\Gamma_{ls}}\frac{\Gamma_{(d-l)s+t}}{\Gamma_{ds+t}}
	\!\!&\to&\displaystyle\left\{\begin{array}{ll}
	(l/d)^{t} ~~~& \substack{d \to \infty\\s \to \infty} \\[6pt]
	t!/d^{t} ~~~& \substack{d \to 1\\s \to 1}
	\end{array}\right. \\
	\!\!\!\!\textrm{Optima:}& \!\!\!\!&x_{*} \in&~ \left\{0,\frac{ls-1}{ds-2},1\right\}
	\!\!&\to&\displaystyle\left\{\begin{array}{ll}
	l/d~~~~~~ & \substack{d \to \infty\\s \to \infty} \\[6pt]
	0 & \substack{d \to 1\\s \to 1}
	\end{array}\right.\!\!\!\!\! \nonumber \\
	\!\!\!\!\textrm{Roots:}& \!\!\!\!&z \in&~ \left\{0,1 \vphantom{\frac{ls-1}{ds-2}}{} \right\} ~,\nonumber
\end{align}
where $\Gamma_{d}=(d-1)!$ is the Gamma function.

In this projector case, the operator rank $l$ and environment dimension $s$ greatly affect the form of the resulting distributions, as per \cref{fig:plot_distribution}. For $l=1$-rank projectors, trivial $s=1$ environments lead to distributions $P(x) \sim (1-x)^{d-2}$ that are monotonically decreasing with $x$ and are peaked at the boundary $x=0$. Conversely, nontrivial $s>1$ environments lead to distributions $P(x) \sim x^{s-1}(1-x)^{(d-1)s-1}$, that are not monotonic with $x$ and are peaked at $0<(s-1)/(ds-2)<1$. Further, when the dimensions $d$ or $s$ are large, these distributions are sharply peaked at $l/d$, with moments as powers of this ratio, representing the proportion of the space that is measured by the operator. Finally, given the shift and scale invariance of the distributions, let us arbitrarily define the $k$-dependent noise scales $1-\gamma^{(k)} = (1-\gamma)^{k}$ and environment dimensions $s^{(k)}=k+1$. Operators become $\Pi \to \Pi_{\gamma}^{(k)} = (1-\gamma^{(k)})\Pi + \gamma^{(k)}[\trace{\Pi}/d]I$, yielding effective global depolarizing-like noisy distributions $P_{\Pi} \to P_{\Pi\gamma}^{(k)}$. Evidently, the rank, dimensions, number of distinct eigenvalues, and their relative eigenspace dimensions, crucially affect the distributions.

\begin{figure}[ht]
	\centering
	\includegraphics[width=\columnwidth]{plot.distribution.N.8.data.pdf}
	\captionsetup{justification=raggedright}
	\caption{Analytical distributions for $l$-rank, $d=2^{8}$-dimensional projectors $\Pi$, with global noise $\gamma$, $P_{\Pi}(x) \!\to\! P_{\Pi\gamma}^{(k)}(x)\!\!=\!\![1/(1\hspace{-1.2pt}-\hspace{-1.2pt}\gamma^{(k)})]P_{\Pi}((x\hspace{-1.2pt}-\hspace{-1.2pt}\gamma^{(k)}l/d)/(1\hspace{-1.2pt}-\hspace{-1.2pt}\gamma^{(k)}))$. For depths $k$ (colors), distributions become peaked around $x=l/d$, as ranks $l=1$ (green-blue) or $l=d/2$ (purple-orange), and environments $s = k+1$, approach $ls \to \infty$, with shifts due to noise $\gamma^{(k)} = 1-(1-\gamma)^{k}$, for $\gamma = 0$ (solid) or $\gamma = 10^{-2}$ (dashed).}
	\label{fig:plot_distribution}
	\vspace{-0.3cm}
\end{figure}

\subsection{Hermitian operators}
Second, let us study properties for general operators,
\begin{align}
	\Pi = \sum_{\xi}\xi I_{\xi}
	\quad : \quad \sigma \leq \xi \leq \lambda~,
\end{align}
with $\#$ distinct eigenvalues $\{\xi,d_{\xi}\}$, and expectation values $\sigma \leq x \leq \lambda$, that follow the polynomial distributions,
\vspace{-13pt}
\begin{align}
{
	\setcounter{equation}{\theequation+1}
	\!\!\!\!\!\!\!
	P_{\Pi}(x) =
	 \left\{\begin{array}{ll}
	 \vspace{-16pt}
	\displaystyle\!\!\sum_{\substack{\xi~\!,~\! l_{\xi} \in [d_{\xi}s]}}\!\!\!\!\pi_{l_{\xi}} ~\sign{\xi-x}~(\xi-x)^{(d-d_{\xi})s+l_{\xi}-1} \!\!\!\!\!& \begin{array}{l}~\\\!\!\!\#>1\\[5pt](\theequation) \end{array} \\[36pt]
	\displaystyle \frac{1}{\lambda - \sigma}\frac{\Gamma_{ds}}{\Gamma_{d_{\sigma}s}\Gamma_{d_{\lambda}s}}~\left(\frac{x-\sigma}{\lambda-\sigma}\right)^{d_{\lambda}s-1}\!\!\left(\frac{\lambda-x}{\lambda-\sigma}\right)^{d_{\sigma}s-1} \!\!\!\!\!\!\!\!\!\!\!\!& \!\!\# = 2 \\[16pt]
	\displaystyle\dirac{x-\lambda} \!\!\!\!\!\!\!\!\!\!\!\!& \!\!\#=1
	\end{array}\right. \nonumber
}
\end{align}
with $t$-order moments for this distribution of,
\begin{align}
	x_{t}
	=&~ \displaystyle\sum_{\substack{\xi~\!,~\! l_{\xi} \in [d_{\xi}s]}} \chi_{l_{\xi},t}~~\xi^{t} ~,
\end{align}
which we show are the complete homogeneous symmetric polynomials in the spectra $\{\xi,d_{\xi}\}$ \cite{krantz1992primer}, with coefficients,
\begin{align}
	\!\!\!
	\pi_{l_{\xi}} =&~ \frac{1}{2}~(-1)^{l_{\xi}}~\frac{\Gamma_{ds}}{\Gamma_{d_{\xi}s-l_{\xi}}\Gamma_{(d-d_{\xi})s+l_{\xi}}} \\
	&~\times~~\sum_{\sum_{\zeta \neq \xi}l_{\zeta}=l_{\xi}} \prod_{\zeta \neq \xi}\binom{d_{\zeta}s+l_{\zeta}-1}{l_{\zeta}}\frac{1}{(\xi-\zeta)^{d_{\zeta}s+l_{\zeta}}} \nonumber \\
	\!\!\!
	\chi_{l_{\xi},t} =&~ ~~~(-1)^{l_{\xi}}~~\frac{\Gamma_{ds}}{\Gamma_{d_{\xi}s-l_{\xi}}}\frac{\Gamma_{d_{\xi}s-l_{\xi}+t}}{\Gamma_{ds+t}} \\
	&~\times~~\! \sum_{\sum_{\zeta \neq \xi}l_{\zeta}=l_{\xi}}~\!\!\!\prod_{\zeta \neq \xi} \binom{d_{\zeta}s+l_{\zeta}-1}{l_{\zeta}} \frac{\xi^{d_{\zeta}s}~\zeta^{l_{\zeta}}}{\left(\xi-\zeta\right)^{d_{\zeta}s+l_{\zeta}}} \! \nonumber ~.
\end{align}

\subsection{Empirical distributions of measurement probabilities}
We now seek to understand the behavior of empirical distributions of expectation values from simulated noisy random quantum circuits, as detailed in \cref{sec:simulations_of_expectation_values_of_operators}, using the library \cite{duschenes2022simulation}, with data available \cite{duschenes2026datadistributions}.

\begin{figure}[h]
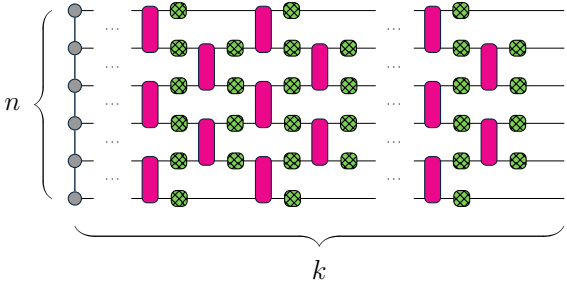

	\tikzfig{0.5}{circuit}
	\captionsetup{justification=raggedright}
	\caption{Circuit with $k$ layers of $n$ qubit brickwork layouts of unitaries (pink) and local noise (green).}
	\label{fig:circuit}
\end{figure}

Here, we sample random quantum channels $\Lambda_{\gamma}^{(k)}$, as per \cref{fig:circuit}, with $k$ layers of one-dimensional brickwork layouts of $n$ qubit, $d = q^{n} ~,~ q = 2$, Haar random nearest-neighbor unitaries $\mathcal{U}$, interspersed with local depolarizing noise $\mathcal{N}_{\gamma}$ with noise scale $0 \leq \gamma \leq q^{2}/(q^{2}-1)$,
\begin{align}
	\mathcal{U} &= \textstyle\prod_{i,j \in [\substack{n\\\textrm{brick}}]} ~U_{ij} ~\!:~\! U_{ij} ~\sim ~\mathcal{U}(q^{2}) \\
	\mathcal{N}_{\gamma} &= \otimes_{i \in [n]}~\mathcal{N}_{\gamma i} ~\!:~\! \mathcal{N}_{\gamma i}(\rho) = (1-\gamma)\rho + \gamma \frac{\trace[i]{\rho}}{q} \otimes I_{i} \nonumber~.
\end{align}
\vspace{-0.5cm}
\vspace{-0.25cm}

\begin{figure}[htp]
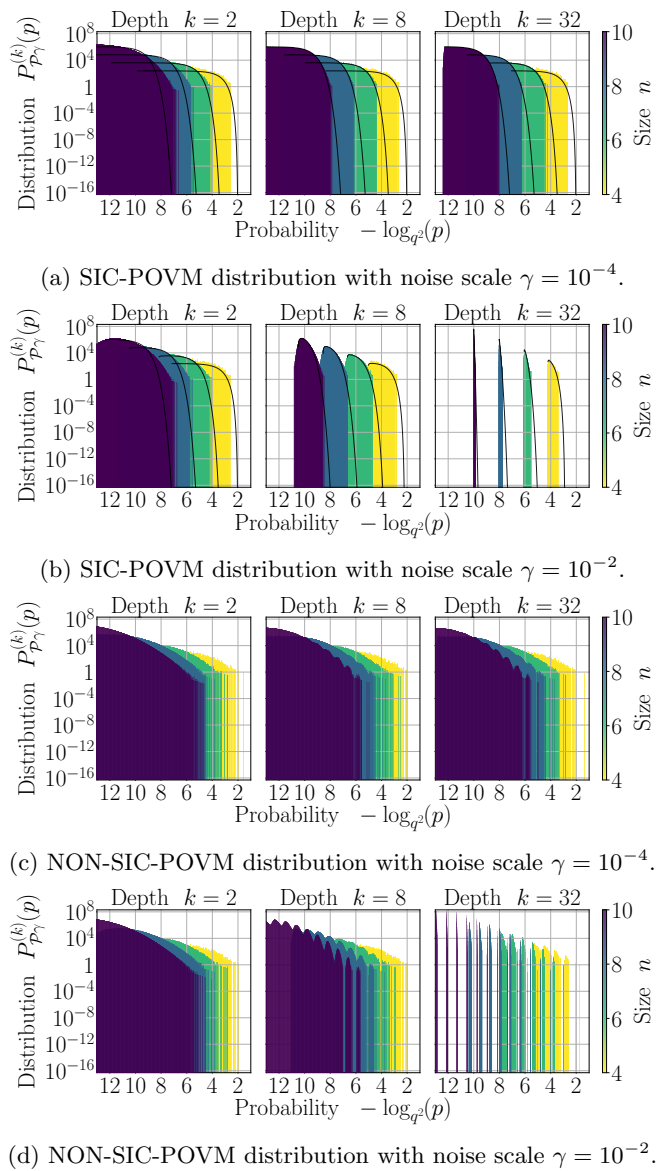

	\centering
	\begin{subfigure}[t]{\columnwidth}
		\centering
		\includegraphics[width=1\columnwidth]{plot.sample.array.M.noise.parameters.N.tetrad.4.pdf}
		\subcaption{SIC-POVM distribution with noise scale $\gamma = 10^{-4}$.}
		\label{fig:plot_probability_sic_nonsic_povm_a}
	\end{subfigure}
	\hfill
	\begin{subfigure}[t]{\columnwidth}
		\centering
		\includegraphics[width=1\columnwidth]{plot.sample.array.M.noise.parameters.N.tetrad.2.pdf}
		\subcaption{SIC-POVM distribution with noise scale $\gamma = 10^{-2}$.}
		\label{fig:plot_probability_sic_nonsic_povm_b}
	\end{subfigure}
	\hfill
	\begin{subfigure}[t]{\columnwidth}
		\centering
		\includegraphics[width=1\columnwidth]{plot.sample.array.M.noise.parameters.N.pauli.4.pdf}
		\subcaption{NON-SIC-POVM distribution with noise scale $\gamma = 10^{-4}$.}
		\label{fig:plot_probability_sic_nonsic_povm_c}
	\end{subfigure}
	\hfill
	\begin{subfigure}[t]{\columnwidth}
		\centering
		\includegraphics[width=1\columnwidth]{plot.sample.array.M.noise.parameters.N.pauli.2.pdf}
		\subcaption{NON-SIC-POVM distribution with noise scale $\gamma = 10^{-2}$.}
		\label{fig:plot_probability_sic_nonsic_povm_d}
	\end{subfigure}
	\captionsetup{justification=raggedright}
	\caption{Empirical SIC-POVM (a,b) and NON-SIC (c,d) probability histograms $\tilde{P}^{(k)}_{\mathcal{P}\gamma}(p)$ for various depths $k$, noise scales $\gamma$, system sizes $n$ (colored histograms), and $m=128$ samples, for Haar random brickwork and local depolarization circuits. Solid lines indicate the fit effective analytical SIC-POVM distributions $P_{\mathcal{P}\gamma}^{(k)}$ for each system size, which generally capture the empirical distributions' peaks, particularly as depth or noise scales increase and the distribution peaks sharpen and tails decrease. As depth and noise scales increase, the empirical NON-SIC-POVM distributions develop multiple sharpened peaks that scale with system size, due to its contributions from $d^{2}$ nonsymmetric measurement operators, each contributing shifted and peaked conditional distributions. Additional plots of SIC and NON-SIC POVM distributions for noise values, $\gamma \in \{0,10^{-4},10^{-3},10^{-2},10^{-1}\}$, are shown in \cref{fig:plot_probability_sic_povm,fig:plot_probability_nonsic_povm} in \cref{sec:simulations_of_expectation_values_of_operators}.}
	\label{fig:plot_probability_sic_nonsic_povm}
\end{figure}

Here, we study sets of positive-operator-valued (POVM) \cite{renes2003symmetric,heinosaari2019random,carrasquilla2019reconstructing,denzler2026simulation,yashin2020minimal,scott2006tight} measurement operators,
\begin{align}
	\mathcal{P} = \{0 \leq \Pi \leq I\}
	\quad : \quad \sum_{\Pi \in \mathcal{P}} \Pi = I~.
\end{align}
Expectation values $0 \leq p \leq 1$ can thus be interpreted as measurement probabilities, with distributions $P_{\mathcal{P}}(p)$ derived from conditional $P_{\Pi}(p)$, given $\Pi \sim P_{\Pi|\mathcal{P}}(\Pi)$,
\begin{align}
	P_{\mathcal{P}}(p) =&~ \sum_{\Pi \in \mathcal{P}}P_{\Pi|\mathcal{P}}(\Pi)~P_{\Pi}(p)
	~.
\end{align}
Examples of such sets of measurement operators include,
\begin{align}
	\!\!\!\!
	\begin{array}{ll}
		\textrm{PVM (projector)} & \Pi = I_{\Pi} \\ 
		\textrm{SIC-POVM (quasiprojector)} & \Pi = \lambda I_{\Pi} \\ 
		\textrm{NON-SIC-POVM (nonprojector)} & \Pi = \sum_{\xi} \xi~I_{\xi} ~. 
	\end{array}
\end{align}
As shown in \cref{fig:plot_probability_sic_nonsic_povm}, we simulate circuits with depths $k \in \{0,2,4,8,16,32\}$, noise scales $\gamma \in \{0,10^{-4},10^{-3},10^{-2},10^{-1}\}$, system sizes $n \in \{4,6,8,10\}$, in $q=2$ qubit initial product pure states, and $m \leq 128$ samples. Operators in $\mathcal{P}$ are uniformly, deterministically measured, $P_{\Pi|\mathcal{P}}(\Pi) = 1/\abs{\mathcal{P}}$, yielding $m\abs{\mathcal{P}}$ samples of measurement probabilities $\{p_{i}\}_{i \in [m\abs{\mathcal{P}}]} \to \{p^{\prime}_{i}\}_{i \in [m^{\prime}]}$, which are further uniformly logarithmically binned into $m^{\prime} = 10^{4}$ samples $p_{i}^{\prime} \in [10^{-20},1]$, resulting in the binned empirical total distributions, $\tilde{P}_{\mathcal{P}}(p) = \sum_{\Pi \in \mathcal{P}}P_{\Pi|\mathcal{P}}(\Pi)\tilde{P}_{\Pi}(p) \to \frac{1}{m^{\prime}}\sum_{i \in [m^{\prime}]}\tilde{P}^{\prime}_{\mathcal{P}}(p|p^{\prime}_{i})$.

As we vary the system parameters $k,\gamma,n,m$, the empirical distribution $\tilde{P}_{\mathcal{P}\gamma}^{(k)}$ of samples $p = \tr({\Pi~\Lambda^{(k)}_{\gamma}(\rho)})$, will converge to a potentially biased distribution $\hat{P}_{\mathcal{P}}$,
\begin{align}
	\!\!\!
	\tilde{P}_{\mathcal{P}\gamma}^{(k)}
	 \to
	\hat{P}_{\mathcal{P}} ~\approx~ P_{\mathcal{P}}
	~\leftrightarrow~ \mathcal{L}_{\mathcal{P}\gamma}^{(k)} ~=~
	\mathcal{L}\left(\tilde{F}^{(k)}_{\mathcal{P}\gamma}, \hat{F}_{\mathcal{P}}\right)
	\to
	0~\!.\!\!\!
\end{align}
In fact, local noisy brickwork systems do not yield exactly globally Haar random mixed states in finite settings \cite{belkin2024approximate}, $\tilde{P}_{\mathcal{P}\gamma}^{(k)} \to \hat{P}_{\mathcal{P}} \neq P_{\mathcal{P}}$, and $\hat{P}_{\mathcal{P}}$ is unknown relative to the known Haar distribution $P_{\mathcal{P}}$. However, such systems are appropriate benchmarks given their experimentally relevant structure \cite{czischek2021simulating,kim2023evidence} and recent analysis \cite{sauliere2026universality,sauliere2025chaotic,deluca2025universality,belkin2024approximate,loio2025correlations}.

To partially account for these biases, we propose an effective distribution $P_{\mathcal{P}\gamma}^{(k)} \approx \hat{P}_{\mathcal{P}}$ of global depolarizing-like noise, $p \to (1-\tilde{\gamma})p + \tilde{\gamma}\trace{\Pi}/d$, with variable effective noise scales $\tilde{\gamma}$ and environment dimensions $\tilde{s}$,
\begin{align}
	P^{(k)}_{\Pi\gamma}(p) \equiv&~ \frac{1}{1-\tilde{\gamma}}P_{\Pi}\left(\frac{p-\tilde{\gamma}\trace{\Pi}/d}{1-\tilde{\gamma}}\right)~.
\end{align}
Given the shift and scale invariance of the distributions for $\Pi$, we are free to shift operators by arbitrary scales $\tilde{\gamma}$, reflecting global depolarization, with arbitrary environment dimensions $\tilde{s}$, reflecting global entanglement with the environment. Such interpretable parameters thus represent effective changes to distributions, given actual local noise. Such a model is inspired by recent bounds on the convergence of locally noisy to globally noisy distributions \cite{dalzell2021random,deshpande2022tight}. This effective model can be fit using empirical data, via minimization of the error between $P^{(k)}_{\Pi\gamma}(p),\tilde{P}^{(k)}_{\Pi\gamma}(p)$, before comparing $F^{(k)}_{\Pi\gamma}(p),\tilde{F}^{(k)}_{\Pi\gamma}(p)$ with the empirical Kolmogorov–Smirnov metric $\tilde{\mathcal{L}}^{(k)}_{\mathcal{P}\gamma} \approx \mathcal{L}_{\mathcal{P}\gamma}^{(k)}$.

First, we simulate SIC-POVM operators, namely $l=1$-rank quasiprojectors with scalings $\lambda = 1/d$, as per \cref{fig:plot_probability_sic_nonsic_povm_a,fig:plot_probability_sic_nonsic_povm_b}. The distributions correspond to those of pure states at low depths or noise scales, and those of depolarized states at large depths and noise scales, with respective smooth decays or sharp peaks at $\sim (l/d)\lambda$.

Second, we simulate NON-SIC-POVM operators, namely $l \leq d$-rank non-projectors, as per \cref{fig:plot_probability_sic_nonsic_povm_c,fig:plot_probability_sic_nonsic_povm_d}. At low depths or noise scales, the distributions are more uniformly spread out than the SIC distributions. At large depths or noise, the distributions become sharply peaked, with a number of peaks that scales with system size. Unlike in the symmetric case where each of the identical contributing conditional distributions have a single common peak, in the nonsymmetric case many of the distinct contributing conditional distributions have distinct peaks. Any differences in spectra of operators across a set directly contribute to multimodality in distributions.

\begin{figure}[htp]
	\centering
	\begin{subfigure}[t]{\columnwidth}
		\centering
		\includegraphics[width=1.03\columnwidth]{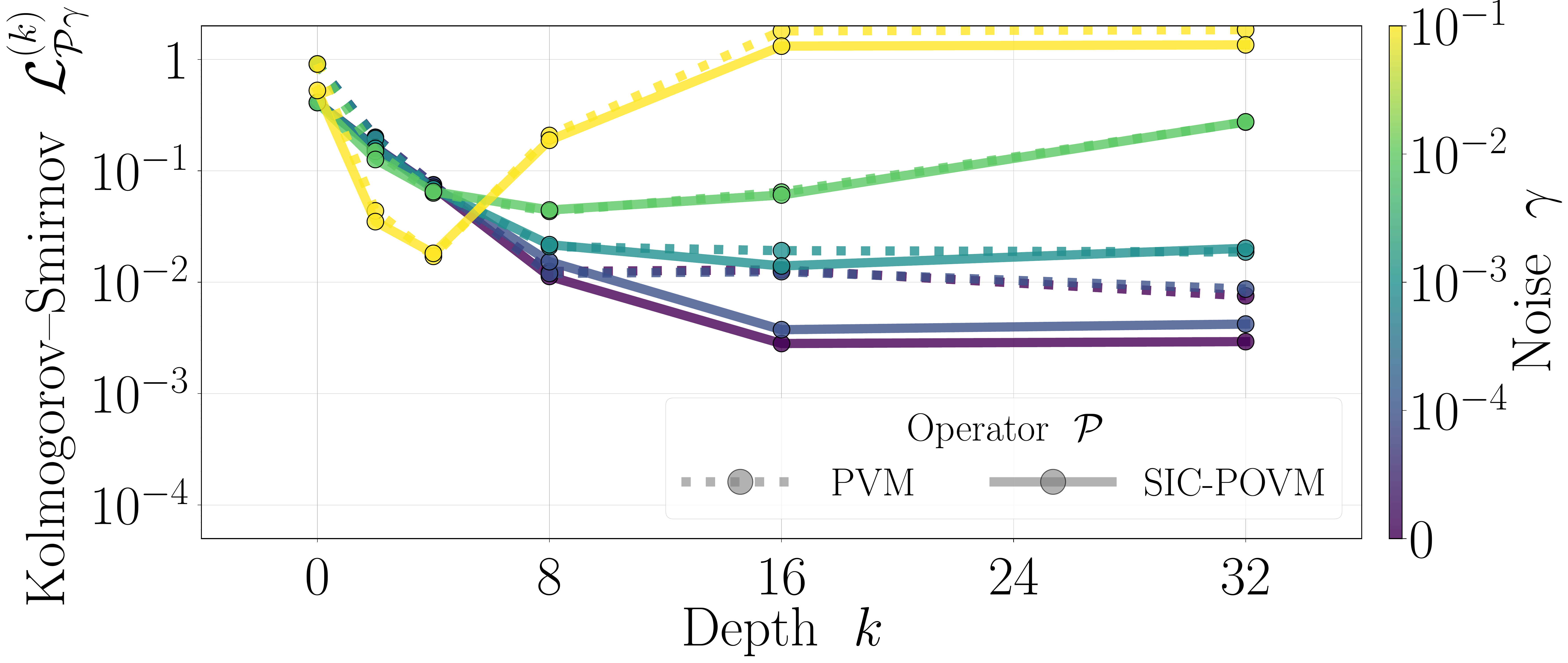}
		\subcaption{Empirical metric $\tilde{\mathcal{L}}^{(k)}_{\mathcal{P}\gamma}$ for size $n=10$.}
		\label{fig:plot_metric_sic_povm_10}
	\end{subfigure}
	\hfill
	\begin{subfigure}[t]{\columnwidth}
		\centering
		\includegraphics[width=\textwidth]{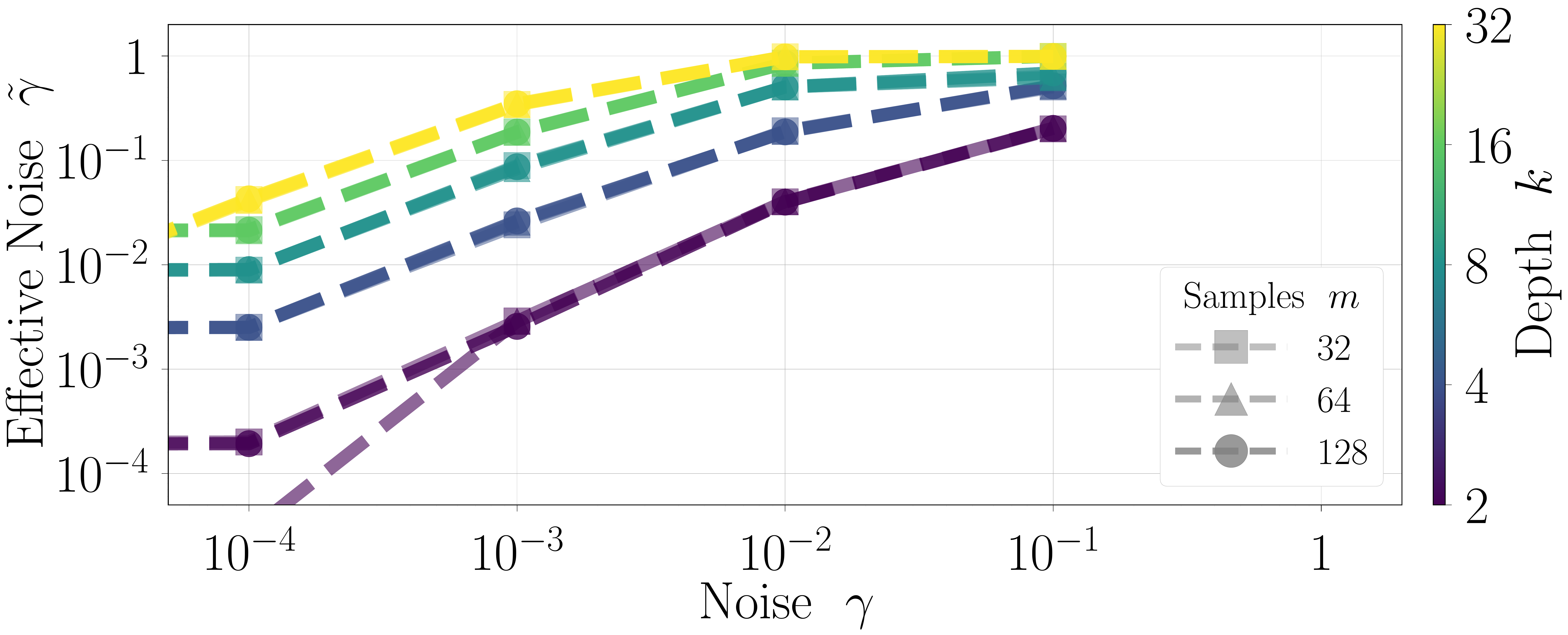}
		\subcaption{Effective noise $\tilde{\gamma}$ versus noise scale $\gamma$.}
		\label{fig:plot_parameter_sic_povm_noise_noise}
	\end{subfigure}
	\hfill
	\begin{subfigure}[t]{\columnwidth}
		\centering
		\includegraphics[width=\textwidth]{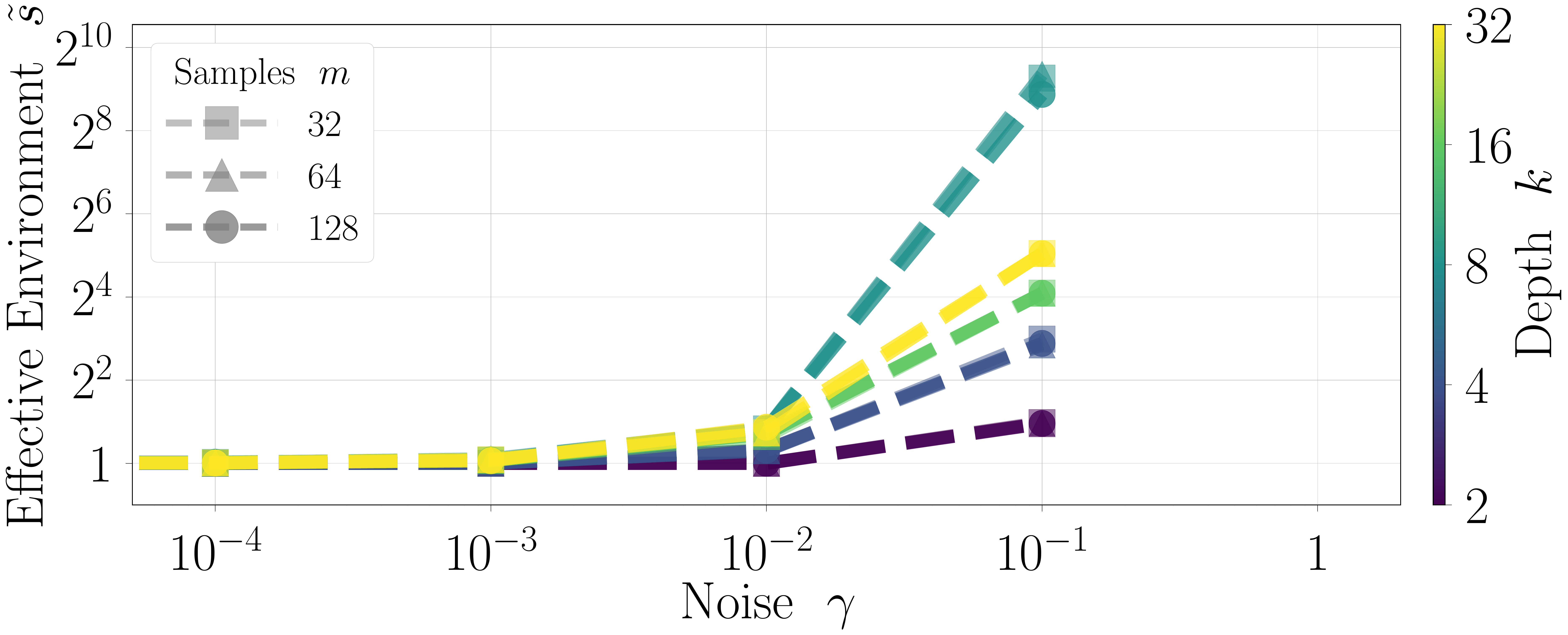}
		\subcaption{Effective environment $\tilde{s}$ versus noise scale $\gamma$.}
		\label{fig:plot_parameter_sic_povm_noise_env}
	\end{subfigure}
	\captionsetup{justification=raggedright}
	\caption{(a): Empirical Kolmogorov–Smirnov metric $\tilde{\mathcal{L}}^{(k)}_{\mathcal{P}\gamma}$ of upper-bounded maximum difference between empirical and analytical cumulative distributions for SIC-POVM and PVM distributions, as a function of depth $k$, for noise scales $\gamma$ (colors), system size $n=10$, and $m=128$ samples, for Haar random brickwork and local depolarization circuits. As depth increases, the metrics converge towards zero, before plateauing due to inherent biases in the effective models. (b,c): Optimized effective noise $\tilde{\gamma}$ and effective environment dimension $\tilde{s}$ as a function of noise scale $\gamma$ and depth $k$ (colors). Additional plots of Kolmogorov–Smirnov metrics and effective model parameters, for system sizes $n \in \{4,6,8,10\}$, and samples $m \in \{32,64,128\}$, are shown in \cref{fig:plot_metric_sic_povm_32_64_128,fig:plot_parameters_sic_povm} in \cref{sec:simulations_of_expectation_values_of_operators}.}
	\label{fig:plot_metric_parameter_sic_povm}
	\vspace{-14pt}
\end{figure}

To quantify these behaviors, we calculate empirical Kolmogorov–Smirnov metrics, as per \cref{fig:plot_metric_parameter_sic_povm}. Such metrics decrease with depth until plateauing for small noise, or increasing for large noise. Such behaviors are primarily attributed to inherent biases in the model at capturing peak versus tails behaviors. There is also potential overfitting of probability densities, which does not necessarily generalize to fitting of cumulative distributions.

Regarding finite-size and finite-sample effects, the metrics converge to smaller values with increased system size $n$, and behave similarly across number of samples $m$. The upper bounded empirical metrics are also non-zero at finite binning, $\tilde{\mathcal{L}} \sim \tbigO{1/m^{\prime}}>0$, leading to inherent bias. Such biases may mask some sample-complexity-dependencies, but do not prevent peaks behaviors from being well described by the effective model. Finally, independent of binning, we may require an exponentially greater number of samples with system size to capture tails information about this distribution on the logarithmic scale. Subsequently, the precise sample complexity and accuracy of the effective model may be masked by being in this relatively low sample regime.

\section{Discussion}\label{sec:discussion}

In this work, we investigate analytical and empirical distributions of expectation values with respect to Haar random states, with an environment traced out. In particular, distributions over sets of non-projective and not-necessarily-symmetric Hermitian operators are studied in depth, demonstrating important differences to previously studied individual projectors \cite{dunkl2011numerical,camposvenuti2013probability,sauliere2025chaotic}.

Using a combinatorics and geometry based approach in \cref{sec:distributions_of_expectation_values_of_operators}, we generalize expressions for expectation value distributions \cite{camposvenuti2013probability}, to the case of mixed states with nontrivial environments, and provide expressions for their moments. Whereas systems with trivial environments or with unit-rank operators have only peaks at the domain boundaries, systems with nontrivial environments or higher-rank operators become sharply, exponentially in system size, peaked at intermediate domain points. The forms of moments are shown to have intimate connections with symmetric polynomials \cite{krantz1992primer}, hypergeometric functions \cite{schlosser2013computer}, and the symmetric subspaces \cite{mele2023introduction} of each of the eigenspaces of the operator. Further, when such distributions are conditional distributions over a set of operators, relative differences in the operator's spectra impose that the total distribution is multiply-peaked, with the number of peaks increasing with system size.

By sampling measurement probabilities given a set of measurement operators from instances of noisy quantum circuits, we confirm the predicted distribution behaviors. Our analytical noiseless expressions subsequently inspire a proposed effective model, with variable effective global noise scale and environment dimension. The effective models are shown to be capable of describing peaks, and less capable of describing tails of the distributions. These behaviors are attributed to a combination of limited samples in these tails, and due to inherent biases in the effective global noise model at capturing local noise behaviors, with its explicit Beta functional form. Potentially obscuring the analysis is also the approximations made involving Kolmogorov–Smirnov upper bounds and binning procedures, which overestimate model errors around sharp peaks in distributions. There also exist tail-sensitive metrics which compute the average total difference in cumulative distributions over the entire domain \cite{stephens1986tests}. Such metrics may be less sensitive to sharply peaked distributions, will more precisely quantify tail behaviors, and are expected to lead to similar conclusions about our overall model's appropriateness.

As discussed in \cref{sec:simulations_of_expectation_values_of_operators}, it is important to assess the validity and interpretability of these effective models. We find the effective noise $\tilde{\gamma}$ and environment dimension $\tilde{s}$ vary with noise scale $\gamma$ and circuit depth $k$ as the approximately polynomial functions, for some threshold $\gamma^{*}$,
\begin{align}
	\begin{array}{l}
	\tilde{\gamma} \sim \bigO{\poly{\gamma}} \\
	\tilde{s} \sim 1 ~+~ \bigO{\poly{\gamma}}\delta_{\gamma > \gamma^{*}}
	\end{array}
	\quad : \quad
	\lim_{\gamma \to 0}\begin{array}{l}\tilde{\gamma} \to 0 \\\tilde{s} \to 1\end{array}
	~.
\end{align}

Curiously, the effective noise scale varies more smoothly with noise scale than the effective environment. The effective environment is also largest at intermediate depths, potentially indicative of where mixed states are maximally Haar random. Although it is not proven that such trends are indicative of global optima of the model, these smooth relationships indicate the optimization is within a physically valid local minimum.

Regarding the universality of our results, other operators could equally be studied using the derived analytical expressions, replacing operators and their spectra $\Pi \to \Lambda^{\dagger}(\Pi)$ with those corresponding with the adjoint action of a channel $\Lambda$. Depending on the fixed points of the channel \cite{duschenes2025moments}, such distributions may be multiply-peaked, and spread out over the logarithmic scale. Further, any non-isotropic noise models may not be able to use the derived shift and scale invariance of the distributions, requiring more thought into appropriate forms of effective models. In particular, the effects of non-unital noise, which typically drive states towards pure states, possibly result in significantly less anti-concentrated, or less uniformly spread out distributions than we observe for our unital depolarizing noise for both SIC and NON-SIC measurement operators \cite{mele2024noise,magni2025anticoncentration}.

We also note that analytical forms of the NON-SIC-POVM measurement operator sets also follow from the above derivations. However, such computations scale at least thrice-exponentially in system size (summations over $\tbigO{\poly{d^{2}}}$ operators $\times$ $\tbigO{\poly{d}}$ eigenvalues $\times$ $\tbigO{\poly{l^{l}}}$ terms per multiplicity-$l$ eigenvalue, per evaluation of $P_{\mathcal{P}}(x)$), and involve exponentially small and large numbers, making them so far numerically infeasible to compare to simulated empirical distributions.

This work sets the stage for further investigations into models for distributions of noisy expectation values. First, our effective models could be supplemented with recent estimates of moments of local noisy systems \cite{sauliere2026universality}, and offer insights into tomography and classical shadow techniques \cite{koh2020classical,jnane2024quantum,huang2020predicting}. Second, moments of such expectation values appear to have deep formal connections to algebraic combinatorics \cite{mele2023introduction,ragone2022representation}, and in principle, quantities such as cross entropy benchmarks \cite{bouland2018on} for our multi-model distributions can be computed.

Finally, given the recent back and forth between quantum and classical methods on simulating noisy random quantum circuits \cite{hangleiter2023computational,zlokapa2023boundaries,aharonov2023polynomial,bouland2018on,pan2022solving,arute2019quantum,boixo2018characterizing,larose2024brief,kim2023evidence,zhu2022quantum}, it should be emphasized that those results are for strictly projective measurement operators, and are typically assumed to converge to unimodal distributions. It will thus be important to assess the computational and sample complexities of our multimodal distributions. For which sets of operators and distributions of quantum states, does any claimed advantage of quantum versus classical methods hold?

\begin{acknowledgments}
	M.D., R.M., and J.C. would like to acknowledge the support of the Natural Sciences and Engineering Research Council of Canada (NSERC) and Compute Canada. Research at the Perimeter Institute is supported in part by the Government of Canada through the Department of Innovation, Science and Economic Development Canada and by the Province of Ontario through the Ministry of Economic Development, Job Creation and Trade. M.D. would also like to acknowledge support from Mike and Ophelia Lazaridis, and companies sponsoring the Vector Institute (\href{https://www.vectorinstitute.ai/partnerships/current-partners/}{https://www.vectorinstitute.ai/partnerships/current-partners/}).
\end{acknowledgments}

\bibliography{main}

\clearpage
\newpage

\onecolumngrid

\appendix

\section{Distributions of expectation values of operators}\label{sec:distributions_of_expectation_values_of_operators}
In these Appendices, we seek to understand the analytical behavior of expectation values of operators, with respect to quantum states. In particular, when the quantum states are randomly distributed, according to some distribution, then the resulting expectation values will also be randomly distributed, according to a yet-to-be-determined distribution. Here, we will extend the contributions of \cite{camposvenuti2013probability} from the case of pure states to the case of mixed states, and provide geometric-based derivations that simplify analyses.

\subsection{Expectation values of operators}
Here, we consider quantum systems described by general $d$-dimension $s$-rank mixed states $\rho$, that are positive, with unit-trace, $\rho \geq 0 ~,~ \trace{\rho} = 1$. Such states can be expressed via purification, in terms of $ds$-dimensional pure states $\psi = \ket{\psi}\bra{\psi}$ in a composite $d$-dimensional system and $s$-dimensional environment space,
\begin{align}
	\rho = (I \otimes \tr[s])(\psi)~,
\end{align}
given the $d$-dimensional identity $I$, and $s$-dimensional partial trace $\tr[s]$. Such states can represented by $d \times s$ complex parameters $\varphi$, via the mapping in terms of elements $ \mu \in [d]~,~ \nu \in [s]$, where $ [n] = \{0,1,\dots,n-1\}$,
\begin{align}
	\varphi
	~ \to ~
	\ket{\psi} = \!\!\!\!\sum_{\mu \in [d]~\!,~\!\nu \in [s]}\!\!\!\! \varphi_{\mu\nu}\ket{\mu\nu}
	~ \to ~
	\rho = \varphi\varphi^{\dagger}~.
\end{align}

We also consider $l$-rank $d$-dimensional Hermitian operators,
\begin{align}
	\Pi = \sum_{\xi} \xi~I_{\xi}~,
\end{align}
with $\#$ number of distinct real eigenvalues $\{\sigma \leq \xi \leq \lambda\}$, with associated $d_{\xi}$-dimensional eigenspaces, with projectors $I_{\xi}$ such that $\sum_{\xi}I_{\xi} = I$. Expectation values $x$ of such operators, with respect to such states, are thus,
\begin{align}
	x = \tau_{\Pi}(\rho) = \trace{\Pi~\rho}~,
\end{align}
which are bounded by the maximum $\lambda$ and minimum $\sigma$ eigenvalues of the operators,
\begin{align}
	\sigma I \leq \Pi \leq \lambda I
	\quad \quad , \quad \quad
	\sigma \leq x \leq \lambda~.
\end{align}
If we assume states $\rho \sim P_{\rho}$ are distributed in terms of their parameters $\varphi \sim P_{\varphi}$ as,
\begin{align}
	P_{\rho}(\rho)
	=&~ \int d\varphi~P_{\varphi}(\varphi)~\dirac{\rho-\varphi\varphi^{\dagger}} ~,
\end{align}
then expectation values $x \sim P_{\Pi}$ are thus randomly distributed, with distributions,
\begin{align}
	P_{\Pi}(x)
	=&~ \int d\rho~P_{\rho}(\rho)~\dirac{x-\tau_{\Pi}(\rho)} = \int d\varphi~P_{\varphi}(\varphi)~\dirac{x-\tau_{\Pi}(\rho)}~.
\end{align}
Here, the complex parameters $\varphi = \alpha + i\beta$ have a measure of,
\begin{align}
	d \varphi = \!\!\!\!\prod_{\mu \in [d]~\!,~\!\nu \in [s]}\!\!\!\!d\alpha_{\mu\nu}~d\beta_{\mu\nu} = d\Omega_{2ds}~d\norm{\varphi} ~\norm{\varphi}^{2ds-1}~,
\end{align}
expressed in terms of $2ds$-dimensional real spherical coordinates, radii $\norm{\varphi}^{2}=\trace{\varphi^{\dagger}\varphi}$, areas $\Omega_{2ds} = 2 {\pi^{ds}}/{\Gamma_{ds}}$, and Gamma functions $\Gamma_{d} = (d-1)!$. For $d,s$-dimensional operators $\Gamma,\Gamma^{\prime}$, such that $\varphi \to \Gamma \varphi \Gamma^{\prime\dagger}$, the measure transforms as $d\varphi \to \abs{\det{\Gamma^{\vphantom{\prime}{}}}}^{2s}\abs{\det{\Gamma^{\prime}}}^{2d}~d\varphi$, and is thus invariant $d\varphi \to d\varphi$ under $d$,$s$-dimensional unitaries $U,V$.

We now choose to make some assumptions about the distribution of states and parameters, with the objective of understanding expectation values with respect to approximately uniformly random states. Given the unitary-invariance of the parameter measure $d\varphi$, we will assume the parameter distribution $P_{\varphi} = P_{\varphi}(\norm{\varphi})$ is also unitarily-invariant and depends strictly on the parameter norm $\norm{\varphi}$. In particular, we choose the parameter distribution to solely be a norm, or trace-preservation constraint,
\begin{align}
	P_{\varphi}(\varphi)~d\varphi ~~=~~ \frac{1}{\frac{1}{2}\Omega_{2ds}} \dirac{\norm{\varphi}^{2}-1}d\varphi~.
\end{align}

\subsection{Moments of expectation values of operators}
While studying distributions of expectation values, it is important to first study the underlying distributions of states which generate such distributions. Insight into its properties allows us to derive expressions for moments of expectation values, without requiring expressions for their distributions.

Here, the unitarily-invariant distribution over $ds$ complex parameters,
\begin{align}
	P_{\varphi}(\varphi)~d\varphi ~~=~~ \frac{1}{\frac{1}{2}\Omega_{2ds}} \dirac{\norm{\varphi}^{2}-1}d\varphi ~~\cong~~ d\psi~,
\end{align}
is in fact isomorphic to the Haar distribution $d\psi$ of uniformly random $ds$-dimensional pure states $\varphi\varphi^{\dagger} \to \psi$, with respect to $ds$-dimensional operators $\Pi \to \Pi \otimes I_{s}$, up to tracing over the $s$-dimensional environment.

Such distributions of $ds$-dimensional states have $t$-order moments in terms of projectors $\mathcal{T}_{t}$ onto the $\binom{ds+t-1}{t}$-dimensional symmetric subspace \cite{mele2023introduction} over $t$-copies of the $ds$-dimensional space,
\begin{align}
	\Psi_{t} =&~ \int d\psi~\psi^{\otimes t} = \frac{1}{\binom{ds+t-1}{t}}\mathcal{T}_{t}~,
\end{align}
and $t$-order moments of expectation values $x = \trace{\psi~\Xi}$ of operators $\Pi \to \Xi = \Pi \otimes I_{s}$ are thus,
\begin{align}
	x_{t} =&~ \int d\psi~\trace{\psi~\Xi}^{t} = \trace{\Psi_{t}~\Xi^{\otimes t}}~.
\end{align}
Finally, the symmetric subspace projector $\mathcal{T}_{t}$ can be expressed in terms of $ds$-dimensional unitary representations $\mathcal{V}$ of the $t$-order permutations $\mathcal{S}_{t}$, and we will denote un-normalized moments of $\Xi$, and traces of powers of $\Xi$ as,
\begin{align}
	\mathcal{T}_{t} = \frac{1}{t!}\sum_{\sigma \in \mathcal{S}_{t}} \mathcal{V}_{\sigma}
	\quad \quad , \quad \quad
	\xi_{t} =&~ \trace{\mathcal{T}_{t}~\Xi^{\otimes t}}
	\quad \quad , \quad \quad
	\zeta_{t} = \trace{\Xi^{t}}
	\quad \quad , \quad \quad
	\vartheta_{t} = \sum_{\sum_{\xi}\sum_{l_{\xi} \in [d_{\xi}]}t_{l_{\xi}} = t} \prod_{\substack{\xi\\l_{\xi} \in [d_{\xi}]}}\xi^{t_{l_{\xi}}}~.
\end{align}

Given these definitions, we can derive closed-form expressions for expectation value moments $\xi_{t}$ in terms of the eigenvalues and multiplicities $\{\xi~,~d_{\xi}\}$ of arbitrary $\Xi$. Such expressions have been derived \cite{camposvenuti2013probability} for non-degenerate $\Xi$ with strictly $d_{\xi}=1$, and are potentially known to the combinatorics community, given several intermediate results, particularly regarding relationships between $\xi_{t},\zeta_{t},\vartheta_{t}$, are known \cite{krantz1992primer}. Our derivations use generating functions,
\begin{align}
	\sum_{t}\xi_{t} ~x^{t}~,
\end{align}
to express $\xi_{t}$ by identifying the $t$-order terms in this series in $x$. Given such generating functions, compositions of functions $F(G(x))$ can be written as series expansions, by collecting terms in series expansions for $F(x)$ and $G(x)$,
\begin{align}
	F(x) = \sum_{t}\frac{1}{t!}~f_{t}~x^{t} ~~ ,&~ ~~ G(x) = \sum_{t}\frac{1}{t}~g_{t}~x^{t}
	\quad \to \quad
	F(G(x)) = \sum_{t}\sum_{l}\sum_{\substack{\sum_{k}l_{k}=l\\\sum_{k}l_{k}k = t}}\frac{1}{l!}\binom{l}{\{l_{k}\}}~f_{l}~\prod_{k}\left(\frac{g_{k}}{k}\right)^{l_{k}}~x^{t}
\end{align}
which is known as the Faà di Bruno formula \cite{krantz1992primer} in terms of the partitions $\{l_{k}\}$ of $(t,l)$ that satisfy the constraints of their sum being $l$ and their sum weighted by $k$ being $t$. Useful examples of such generating functions are,
\begin{align}
	F(x) =&~ e^{x} = \sum_{t}\frac{1}{t!}~x^{t} \to f_{t} = 1
	\quad , \quad
	F(x) = \frac{1}{(1-x)^{l}} = \sum_{t}\binom{l+t-1}{t}~x^{t} \to f_{t} = \frac{\Gamma_{l+t}}{\Gamma_{l}} \\
	G(x) =&~ \log{\left(\frac{1}{1-x}\right)} = \sum_{l}\frac{1}{t}~x^{t} \to g_{t} = 1
	\quad \quad , \quad \quad
	\binom{l}{\{l_{\xi}\}} = \frac{l!}{\prod_{\xi}l_{\xi}!}
	\quad \quad , \quad \quad
	\Gamma_{l}=(l-1)! ~.
\end{align}

Partitions $\{l_{k}\}$ of $(t,l)$ are particularly relevant to our derivations. In fact, summations over $t$-order permutations $\mathcal{S}_{t}$ can be described as summations over partitions, with a partition identically describing the cycle structure $\{l_{k}\}$ of a permutation, given permutations with $l \leq t$ cycles have $l_{k}$ number of $k$-length cycles,
\begin{align}
	\frac{1}{t!}\sum_{\sigma \in \mathcal{S}_{t}} ~=~ \frac{1}{l!}~\sum_{l}\sum_{\substack{\sum_{k}l_{k}=l\\\sum_{k}l_{k}k = t}}\binom{l}{\{l_{k}\}}\prod_{k}\frac{1}{k^{l_{k}}}~.
\end{align}
There are $\binom{t}{\{l_{k}k\}}$ ways of choosing such partitions, times $(l_{k}k)!$ ways of placing the elements within the $l_{k}$ $k$-length cycles, less the $l_{k}!$ ways of ordering the $k$-length cycles, and less $k^{l_{k}}$ ways of choosing the representative first element in each $k$-length cycle. We also can replace the $t!$ factors with $l!$, given $t!$ occurs in the numerator and denominator.

To derive the moments $\xi_{t}$, we note that traces of operators with permutations $\sigma \in \mathcal{S}_{t}$, with cycle structure described by the partition $\{l_{k}\}$, equals products of powers of traces of such operators $\zeta_{t}$, allowing us to relate $\xi_{t}$ and $\zeta_{t}$,
\begin{align}
	\trace{\mathcal{V}_{\sigma}~\Xi^{\otimes t}} =&~ \prod_{k}\trace{\Xi^{k}}^{l_{k}} = \prod_{k}\zeta_{k}^{l_{k}}
	\quad \quad \to \quad \quad
	\xi_{t} = \frac{1}{t!}~\sum_{l}\sum_{\substack{\sum_{k}l_{k}=l\\\sum_{k}l_{k}k = t}}\binom{l}{\{l_{k}\}}\prod_{k}\left(\frac{\zeta_{k}}{k}\right)^{l_{k}}~.
\end{align}
Further, given $\zeta_{k}$, $\vartheta_{t}$ are functions of the spectra $\{\xi,d_{\xi}\}$, we have the relationships to distinct generating functions,
\begin{align}
	\zeta_{k} = \sum_{\xi}d_{\xi}\xi^{k}
	\quad \quad \to&~ \quad \quad
	\sum_{k}\frac{1}{k}\zeta_{k}~x^{k} = \sum_{\xi}d_{\xi}\sum_{k}\frac{1}{k}\left(\xi x\right)^{k} = \sum_{\xi}\log{\left(\frac{1}{1-\xi x}\right)^{d_{\xi}}} \\
	\vartheta_{t} = \sum_{\sum_{\xi}\sum_{l_{\xi} \in [d_{\xi}]}t_{l_{\xi}} = t} \prod_{\substack{\xi\\l_{\xi} \in [d_{\xi}]}}\xi^{t_{l_{\xi}}}
	\quad \quad \to&~ \quad \quad
	\sum_{t} \vartheta_{t}~x^{t} = \sum_{t}\sum_{\sum_{\xi}\sum_{l_{\xi} \in [d_{\xi}]}t_{l_{\xi}} = t}\prod_{\xi}\left(\xi x\right)^{t_{l_{\xi}}} = \prod_{\xi}\left(\frac{1}{1-\xi x}\right)^{d_{\xi}}
	~.
\end{align}

Such expansions of permutations in terms of partitions for $\xi_{t}$, are reminiscent of the expansions of compositions of generating functions, namely, the composition of the exponential function with the generating function of $\zeta_{k}$,
\begin{align}
	\!\!\!
	\sum_{t}\xi_{t}~x^{t} = e^{\sum_{k}\frac{1}{k}\zeta_{k}~x^{k}} = \prod_{\xi}\left(\frac{1}{1-\xi x}\right)^{d_{\xi}} = \frac{1}{\det{I - x\Xi}} = \sum_{t} \vartheta_{t}~x^{t}~\!.\!\!
\end{align}

Therefore the $t$-order moments $\xi_{t}$ of operators $\Xi$ with respect to the symmetric subspace projector $\mathcal{T}_{t}$ are in fact the {complete homogeneous symmetric polynomials \cite{krantz1992primer} in the spectra $\{\xi,d_{\xi}\}$, generated by the determinant of $I - x\Xi$,
\begin{align}
	\xi_{t}
	~=&~~ \vartheta_{t}
	~=~ \sum_{\sum_{\xi}\sum_{l_{\xi} \in [d_{\xi}]}t_{l_{\xi}} = t} \prod_{\substack{\xi\\l_{\xi} \in [d_{\xi}]}}\xi^{t_{l_{\xi}}}
	~=~ \frac{1}{t!}~\partial_{x}^{t}~\frac{1}{\det{I - x\Xi}}\vert_{x=0}~.
\end{align}

Finally, we can derive a closed-form expression for such moments $\xi_{t}$, using partial fraction decompositions,
\begin{align}
	\frac{1}{\prod_{\xi}(x-\xi)^{d_{\xi}}} =&~ \sum_{\xi}\sum_{l_{\xi} \in [d_{\xi}]}\frac{f_{l_{\xi}}}{(x-\xi)^{ds-l_{\xi}}}
	\quad \quad : \quad \quad
	f_{l_{\xi}} = \frac{1}{l_{\xi}!}\partial_{x}^{l_{\xi}}\frac{1}{\prod_{\zeta \neq \xi}(x-\zeta)^{d_{\zeta}}}\vert_{x=\xi}~,
\end{align}
where derivatives of products of functions $\{f_{\xi}(x)\}$ and of reciprocal polynomials $1/(x-\xi)^{k}$ can be expanded as,
\begin{align}
	\partial_{x}^{l}\prod_{\xi}f_{\xi}(x) = \sum_{\sum_{\xi}l_{\xi}=l} \binom{l}{\{l_{\xi}\}}\prod_{\xi}\partial_{x}^{l_{\xi}}f_{\xi}(x)
	\quad ,&~ \quad
	\partial_{x}^{l}\frac{1}{(x-\xi)^{k}} = (-1)^{l}~\frac{\Gamma_{k+l}}{\Gamma_{k}}~\frac{1}{(x-\xi)^{k+l}}~.
\end{align}
Therefore an expression for the determinant of $I - x\Xi$ is,
\begin{align}
	\prod_{\xi}\frac{1}{\left(1-\xi x\right)^{d_{\xi}}}
	=&~ \sum_{\xi}\sum_{l_{\xi} \in [d_{\xi}]} (-1)^{l_{\xi}} ~\xi^{ds-d_{\xi}}~\sum_{\sum_{\zeta \neq \xi}l_{\zeta}=l_{\xi}}\prod_{\zeta \neq \xi} \binom{d_{\zeta}+l_{\zeta}-1}{l_{\zeta}} \frac{\zeta^{l_{\zeta}}}{\left(\xi-\zeta\right)^{d_{\zeta}+l_{\zeta}}}~\frac{1}{\left(1-\xi x\right)^{d_{\xi}-l_{\xi}}}~,
\end{align}
and expanding the reciprocal $x$-dependent terms, we can match the $t$-order term with the $t$-order moments,
\begin{align}
	\xi_{t}
	=&~ \sum_{\xi}\sum_{l_{\xi} \in [d_{\xi}]} (-1)^{l_{\xi}}~\binom{d_{\xi}-l_{\xi}+t-1}{t}~\sum_{\sum_{\zeta \neq \xi}l_{\zeta}=l_{\xi}}\prod_{\zeta \neq \xi} \binom{d_{\zeta}+l_{\zeta}-1}{l_{\zeta}} \frac{\zeta^{l_{\zeta}}}{\left(\xi-\zeta\right)^{d_{\zeta}+l_{\zeta}}}~\xi^{ds-d_{\xi}+t}~.
\end{align}
By using combinatorics we avoid any contour integration used in \cite{camposvenuti2013probability}, simplifying our analysis to derive expressions for the $t$-order moments of $d$-dimensional operators $\Pi$ with spectra $\{\xi,d_{\xi}\}$ in $s$-dimensional environments,
\begin{align}
	x_{t} =&~ \displaystyle\sum_{\substack{\xi}}\sum_{\substack{l_{\xi} \in [d_{\xi}s]}} \chi_{l_{\xi},t}~\xi^{t}~,
\end{align}
given the spectrum-dependent coefficients,
\begin{align}
	\chi_{l_{\xi},t} = (-1)^{l_{\xi}}~\frac{\Gamma_{ds}}{\Gamma_{d_{\xi}s-l_{\xi}}}\frac{\Gamma_{d_{\xi}s-l_{\xi}+t}}{\Gamma_{ds+t}}~\!\!\!\!\displaystyle\sum_{\sum_{\zeta \neq \xi}l_{\zeta}=l_{\xi}}\!\prod_{\zeta \neq \xi} &\binom{d_{\zeta}s+l_{\zeta}-1}{l_{\zeta}} \frac{\xi^{d_{\zeta}s}~\zeta^{l_{\zeta}}}{\left(\xi-\zeta\right)^{d_{\zeta}s+l_{\zeta}}} ~.
\end{align}

\newpage
\subsection{Distributions of expectation values of operators}

Here, we consider expectation values $x = \trace{\rho~\Pi} = \norm{\Gamma\varphi}^{2}$, with respect to $l$-rank operators $\Pi = \Gamma^{\dagger}\Gamma$, with $\#$ distinct eigenvalues $\{\sigma \leq \xi \leq \lambda\}$, and $s$-rank states $\rho = \varphi\varphi^{\dagger}$. Here the parameters $\varphi$ are distributed according to the unitarily-invariant distribution of Haar random $ds$-dimensional pure states,
\begin{align}
	\varphi \sim P_{\varphi} \propto \dirac{\norm{\varphi}^{2}-1}
	\quad \to \quad
	x \sim P_{\Pi} \propto \int d\varphi~ \dirac{\norm{\varphi}^{2}-1}~\dirac{\norm{\Gamma\varphi}^{2}-x}~.
\end{align}

From the unitary invariance that defines the expectation value distributions, given any scalars $\zeta,\varsigma$ such that operators are transformed as $\Pi \to (\Pi - \varsigma I)/\zeta$, then the distributions exhibit shift and scale invariance,
\begin{align}
	P_{\Pi}(x) =&~ \frac{1}{\abs{\zeta}}P_{\frac{\Pi - \varsigma I}{\zeta}}\left(\frac{x-\varsigma}{\zeta}\right)
	\quad \quad : \quad \quad
	\sigma I \leq \Pi \leq \lambda I
	\quad \quad , \quad \quad
	\zeta\sigma + \varsigma \leq x \leq \zeta\lambda + \varsigma~.
\end{align}
As such, without loss of generality, it is convenient to normalize the operators as,
\begin{align}
	\Pi \to \frac{\Pi - \sigma I}{\lambda - \sigma}
	\quad , \quad
	x \to \frac{x-\sigma}{\lambda-\sigma}~,
\end{align}
which are bounded by their eigenvalues,
\begin{align}
	0 \leq \Pi \leq I
	\quad \quad , \quad \quad
	0 \leq x \leq 1~.
\end{align}
The normalized operators $\Pi \equiv \Gamma^{\dagger}\Gamma \geq 0$ are thus positive, $l$-rank, with $\#$ number of distinct non-negative eigenvalues. The image of the normalized $\Pi$ thus induces $l,d-l$-dimensional subspaces with associated $l,d-l$-rank projectors $I_{\Pi},I-I_{\Pi}$, and it is often convenient to partition the parameters $\varphi$ into associated $l,d-l$-dimensional components,
\begin{align}
	\varphi \to \varphi \oplus \bar{\varphi} \quad , \quad d\varphi \to d\varphi~d\bar{\varphi}~.
\end{align}
Such normalizations and partitionings enforce that the positive normalized $l<d$-rank operators $\Pi$ always have a nontrivial $l$-dimensional image where $\Pi$ acts on $\varphi$, and a nontrivial $d-l$-dimensional kernel where $\Pi$ acts $\bar{\varphi}$. Such a bi-partitioned basis thus simplifies analysis by straightforwardly satisfying the two distribution Dirac-delta functions constraints related to trace-preservation $\norm{\varphi\oplus\bar{\varphi}}^{2} = 1$ and expectation value-preservation $\norm{\Gamma\varphi}^{2}=x$.~\\

The distribution of $x \sim P_{\Pi}(x)$ for $\sigma \leq x \leq \lambda$, thus depends on the number of eigenvalues of $\#$ of $\Pi$.~\\

For $\#=1$-eigenvalue operators $\Pi = I$,
\begin{align}
	P_{\Pi}(x)
	=&~ \int d\varphi ~P_{\varphi}(\varphi)~\dirac{\pi_{\Pi}(\varphi)-x} \\
	=&~ \int d\varphi ~P_{\varphi}(\varphi)~\dirac{\norm{\varphi}^{2}-x} \\
	=&~ \frac{1}{\frac{1}{2}\Omega_{2ds}}\int d\varphi ~\dirac{\norm{\varphi}^{2}-1}~\dirac{\norm{\varphi}^{2}-x}\\
	=&~ \dirac{x-1}~.
\end{align}

For $\#=2$-eigenvalue operators $\Pi = I_{\Pi}$,
\begin{align}
	P_{\Pi}(x)
	=&~ \int d\varphi ~P_{\varphi}(\varphi)~\dirac{\pi_{\Pi}(\varphi)-x} \\
	=&~ \int d\varphi~d\bar{\varphi} ~P_{\varphi\bar{\varphi}}(\varphi,\bar{\varphi})~\dirac{\norm{\varphi}^{2}-x} \\
	=&~ \frac{1}{\frac{1}{2}\Omega_{2ds}}\int d\varphi~d\bar{\varphi} ~\dirac{\norm{\varphi}^{2}+\norm{\bar{\varphi}}^{2}-1}~\dirac{\norm{\varphi}^{2}-x}\\
	=&~ \frac{\frac{1}{2}\Omega_{2ls}~\frac{1}{2}\Omega_{2(d-l)s}}{\frac{1}{2}\Omega_{2ds}}~x^{ls-1}~\frac{1}{\frac{1}{2}\Omega_{2ls}}\int d\varphi ~\dirac{\norm{\varphi}^{2}-1}\left(1 - x\norm{\varphi}^{2}\right)^{(d-l)s-1}\\
	=&~ \frac{\Gamma_{ds}}{\Gamma_{ls}~\Gamma_{(d-l)s}}~x^{ls-1}~\left(1 - x\right)^{(d-l)s-1} ~.
\end{align}

For $\#>1$-eigenvalue operators $\Pi = \Gamma^{\dagger}\Gamma$,
\begin{align}
	P_{\Pi}(x)
	=&~ \int d\varphi ~P_{\varphi}(\varphi)~\dirac{\norm{\Gamma\varphi}^{2}-x} \\
	=&~ \frac{1}{\frac{1}{2}\Omega_{2ds}}\int d\varphi ~\dirac{\norm{\varphi}^{2}-1} \dirac{\norm{\Gamma\varphi}^{2}-x} \\
	=&~ \frac{1}{\frac{1}{2}\Omega_{2ds}}\int d\varphi~d\bar{\varphi} ~\dirac{\norm{\varphi}^{2}+\norm{\bar{\varphi}}^{2}-1}\dirac{\norm{\Gamma\varphi}^{2}-x} \\
	=&~ \frac{\frac{1}{2}\Omega_{2(d-l)s}}{\frac{1}{2}\Omega_{2ds}}\int d\varphi~\left(1-\norm{\varphi}^{2}\right)^{(d-l)s-1} ~\dirac{\norm{\Gamma\varphi}^{2}-x} \\
	=&~ \frac{\frac{1}{2}\Omega_{2ls}\frac{1}{2}\Omega_{2(d-l)s}}{\frac{1}{2}\Omega_{2ds}}\frac{1}{\abs{\det{\Gamma}}^{2s}}~x^{ls-1}~\frac{1}{\frac{1}{2}\Omega_{2ls}}\int d\varphi~\dirac{\norm{\varphi}^{2}-1}\left(1-x\norm{\Gamma^{-1}\varphi}^{2}\right)^{(d-l)s-1} \\
	=&~ \frac{\frac{1}{2}\Omega_{2ls}\frac{1}{2}\Omega_{2(d-l)s}}{\frac{1}{2}\Omega_{2ds}}\frac{1}{{\det{\Pi}}^{s}}~x^{ls-1}~\frac{1}{\frac{1}{2}\Omega_{2ls}}\int d\varphi~\dirac{\norm{\varphi}^{2}-1}\trace{\varphi\varphi^{\dagger ~\! \otimes (d-l)s-1}~\left(I_{\Pi}-x\Pi^{-1}\right)^{\otimes (d-l)s-1}} \\
	=&~ \frac{\frac{1}{2}\Omega_{2ls}\frac{1}{2}\Omega_{2(d-l)s}}{\frac{1}{2}\Omega_{2ds}}\frac{1}{{\det{\Pi}}^{s}}~x^{ls-1}~\frac{1}{\trace{\mathcal{T}_{t}}}~\trace{\mathcal{T}_{t}~\Delta(x)^{\otimes t}} ~,
\end{align}
where $t = (d-l)s-1$, $q=ls$, $\Delta(x) = \left(I_{\Pi}-x\Pi^{-1}\right) \otimes I$, and $\mathcal{T}_{t}$ is the projector onto the $\binom{q+t-1}{t}$-dimensional symmetric subspace of the $t$-copies of the $q$-dimensional space. Using our expressions for $t$-order moments,
\begin{align}
	P_{\Pi}(x) =&~ \sum_{\xi \neq 0}\sum_{l_{\xi} \in [d_{\xi}s]}\pi_{l_{\xi}}(x)~\frac{x^{d_{\xi}s-l_{\xi}-1}(\xi-x)^{(d-d_{\xi})s-1}}{\xi^{(d-l)s+d_{\xi}s-l_{\xi}-1}} ~,
\end{align}
with spectrum and $x$-dependent coefficients,
\begin{align}
	\pi_{l_{\xi}}(x) =&~ (-1)^{l_{\xi}}\frac{\Gamma_{ds}}{\Gamma_{ls}\Gamma_{(d-l)s}}\frac{\Gamma_{ls}}{\Gamma_{d_{\xi}s-l_{\xi}}}\frac{\Gamma_{(d-l)s+d_{\xi}s-l_{\xi}-1}}{\Gamma_{ds-1}}\sum_{\sum_{\zeta\neq0,\xi}l_{\zeta}=l_{\xi}}\prod_{\zeta \neq 0,\xi}\binom{d_{\zeta}s+l_{\zeta}-1}{l_{\zeta}}\frac{(\zeta-x)^{l_{\zeta}}}{(\xi-\zeta)^{d_{\zeta}s+l_{\zeta}}}~.
\end{align}

Alternatively, we can derive simpler expressions, by extending previous derivations using contour integration \cite{camposvenuti2013probability}, to the case of nontrivial environments $s>1$.

For $\#>1$-eigenvalue operators $\Pi = \Gamma^{\dagger}\Gamma$,
\begin{align}
	\!
	P_{\Pi}(x)
	=&~ \int d\varphi ~P_{\varphi}(\varphi)~\dirac{\pi_{\Pi}(\varphi)-x} \\
	=&~ \frac{1}{\frac{1}{2}\Omega_{2ds}}\int d\varphi ~\dirac{\norm{\varphi}^{2}-1}~\dirac{\norm{\Gamma\varphi}^{2}-x} \\
	=&~ \frac{1}{\frac{1}{2}\Omega_{2ds}}\frac{1}{2\pi2\pi}\int du~dv~e^{-i(xu+v)}\int d\varphi ~e^{i~\!\trace{\varphi^{\dagger}(vI+u\Pi)\varphi}} \\
	=&~ \frac{i^{ds}\pi^{ds}}{\frac{1}{2}\Omega_{2ds}}\frac{1}{2\pi2\pi}\int du~e^{-ixu}\int dv~\frac{e^{-iv}}{\prod_{\xi}(v+u\xi)^{d_{\xi}s}}\\
	=&~ \sum_{\substack{\xi}}\sum_{\substack{l_{\xi} \in [d_{\xi}s]}}\sum_{\sum_{\zeta \neq \xi}l_{\zeta}=l_{\xi}}(-1)^{l_{\xi}}~\frac{\Gamma_{ds}}{\Gamma_{d_{\xi}s-l_{\xi}}}\prod_{\zeta \neq \xi}\binom{d_{\zeta}s+l_{\zeta}-1}{l_{\zeta}}\frac{1}{(\xi-\zeta)^{d_{\zeta}s+l_{\zeta}}} ~\int \frac{du}{2\pi}~\frac{e^{iu(\xi-x)}}{\left(iu\right)^{(d-d_{\xi})s+l_{\xi}}} \\
	P_{\Pi}(x)
	=&~ \sum_{\substack{\xi}}\sum_{\substack{l_{\xi} \in [d_{\xi}s]}}\pi_{l_{\xi}} ~\sign{\xi-x}~(\xi-x)^{(d-d_{\xi})s+l_{\xi}-1} ~,
\end{align}
with cumulative distributions,
\begin{align}
	F_{\Pi}(x)
	=&~ \sum_{\substack{\xi}}\sum_{\substack{l_{\xi} \in [d_{\xi}s]}}\pi_{l_{\xi}} ~\left(-\sign{\xi-x}~(\xi-x)^{(d-d_{\xi})s+l_{\xi}} ~+~ (\xi-\sigma)^{(d-d_{\xi})s+l_{\xi}}\right) ~,
\end{align}
given the spectrum-dependent coefficients,
\begin{align}
	\pi_{l_{\xi}} =&~ \frac{1}{2}(-1)^{l_{\xi}}~\frac{\Gamma_{ds}}{\Gamma_{d_{\xi}s-l_{\xi}}\Gamma_{(d-d_{\xi})s+l_{\xi}}}\sum_{\sum_{\zeta\neq\xi}l_{\zeta}=l_{\xi}}\prod_{\zeta \neq \xi}\binom{d_{\zeta}s+l_{\zeta}-1}{l_{\zeta}}\frac{1}{(\xi-\zeta)^{d_{\zeta}s+l_{\zeta}}} ~.
\end{align}

To derive expressions for these distributions, we use several identities, including for Dirac delta functions,
\begin{align}
	\delta(\xi) =&~ \frac{1}{2\pi}\int dx~e^{i\xi x}~,
\end{align}
and for $ds$-dimensional complex Gaussian integrals with respect to operators $\Pi$ with spectra $\{\xi,d_{\xi}\}$,
\begin{align}
	\int d^{2ds} \varphi~ e^{i~\!\trace{\varphi^{\dagger}\Pi\varphi}} =&~ \frac{\left(i\pi\right)^{ds}}{\abs{\det{\Pi}}^{s}}
	\quad \quad , \quad \quad
	\det{xI + \Pi} = \prod_{\xi}\left(x+\xi\right)^{d_{\xi}}
	\quad \quad , \quad \quad
	\binom{d}{\{d_{\xi}\}} = \frac{d!}{\prod_{\xi}d_{\xi}!}	~.
\end{align}
Integrals can be evaluated via contour integration of products of functions,
\begin{align}
	\int dx~\frac{1}{\prod_{\xi}(x-\xi)^{d_{\xi}}} =&~ i2\pi\sum_{\xi}\sign{\xi}~\frac{1}{\Gamma_{d_{\xi}}}~\partial_{x}^{d_{\xi}-1} \left.\frac{1}{\prod_{\zeta \neq \xi}(x-\zeta)^{d_{\zeta}}} \right\vert_{x=\xi}~,
\end{align}
with the following identities for various derivatives of functions $f,g,\{f_{\xi}\}$,
\begin{align}
	\partial^{l}fg = \sum_{k}^{l}\binom{l}{k}~\partial^{k}f~\partial^{l-k}g
	\quad \quad , \quad \quad
	\partial^{l}\prod_{\xi}f_{\xi} = \sum_{\sum_{\xi}l_{\xi}=l} \binom{l}{\{l_{\xi}\}}\prod_{\xi}\partial^{l_{\xi}}f_{\xi}
	~,
\end{align}
\begin{align}
	\partial_{x}^{l}~\frac{1}{(x-\xi)^{k}} = (-1)^{l}~\frac{\Gamma_{k+l}}{\Gamma_{k}}~\frac{1}{(x-\xi)^{k+l}}
	\quad \quad , \quad \quad
	\partial_{x}^{l}~e^{\xi x} = \xi^{l}~e^{\xi x}
	~,
\end{align}
and therefore we have,
\begin{align}
	\partial_{x}^{d_{\xi}-1}~\frac{e^{zx}}{\prod_{\zeta \neq \xi}(x-\zeta)^{d_{\zeta}}}
	=&~
	\sum_{\substack{l_{\xi} \in [d_{\xi}]}}\sum_{\sum_{\zeta\neq\xi}l_{\zeta}=l_{\xi}}(-1)^{l_{\xi}}~z^{d_{\xi}-l_{\xi}-1}~\frac{\Gamma_{d_{\xi}}}{\Gamma_{d_{\xi}-l_{\xi}}} \prod_{\zeta \neq \xi}\binom{d_{\zeta}+l_{\zeta}-1}{l_{\zeta}}\frac{1}{(x-\zeta)^{d_{\zeta}+l_{\zeta}}}~e^{zx} ~.
\end{align}
Finally, we note that integrals of sign functions are,
\begin{align}
	\int_{\alpha}^{\beta}dx~\sign{\gamma-x}\left(\gamma-x\right)^{k-1} =&~ \frac{1}{k}\left(-\sign{\gamma-\beta}\left(\gamma-\beta\right)^{k} ~+~ \left(\gamma-\alpha\right)^{k}\right)~.
\end{align}

The resulting distribution and moment expressions, in particular their coefficients $\pi_{l_{\xi}},\chi_{l_{\xi},t}$, have remarkable similarities and differences, however do not appear to immediately follow from each other. The forms of such expressions evidently have deep connections to algebraic combinatorics \cite{krantz1992primer}, and the distributions appear to potentially have connections to hypergeometric functions \cite{schlosser2013computer}.

\clearpage
\newpage

\section{Simulations of expectation values of operators} \label{sec:simulations_of_expectation_values_of_operators}
In these Appendices, we describe implementation details of numerically simulating distributions of expectation values of operators. In particular, we discuss the simulated sets of measurement operators, numerical representations of empirical distributions, perform studies of effects of number of samples on these studies, and finally investigate and interpret the properties of the fit effective analytical model parameters. All numerical experiments are performed using a custom Jax-based quantum circuit simulator \cite{duschenes2022simulation}, and data is available \cite{duschenes2026datadistributions}.

\subsection{Sets of measurement operators}
In this work, we consider expectation values, namely measurement probabilities $p$ resulting from sets of positive-operator-valued (POVM) measurement operators $\Pi \in \mathcal{P}$. Such sets of operators may be distinguished by symmetries, namely whether the properties of an operator from the set are independent of the operator itself.

\subsubsection{Local POVM measurement operators}
Regarding the specific operators studied in this work, for numerical efficiency during simulations, we will consider tensor-products of $n$, $q ~:~ d = q^{n}$-dimensional local operators $\Pi \to \Pi = \otimes_{i \in [n]}\Pi_{i}$. Each local $\Pi_{i} \in \mathcal{P}_{i}$ will be from the same set of operators $\mathcal{P}_{i} = \mathcal{P}$, thus $\mathcal{P} \to \mathcal{P}^{\otimes n}$. We will consider both symmetric and nonsymmetric sets of operators.

First, we consider the set of $\abs{\mathcal{P}^{\textrm{PVM}}} = d$ symmetric, local, orthogonal, and non-informationally-complete Projector PVM measurement operators, in terms of orthogonal basis states $\ket{\mu}$, $\mu \in [d]$,
\begin{align}
	\Pi_{\mu}^{\textrm{PVM}} =&~ \otimes_{i \in [n]}\psi_{\mu_{i}}^{\textrm{PVM}} \\
	\ket{\psi}_{\mu_{i}}^{\textrm{PVM}} =&~ \ket{\mu_{i}} ~. \nonumber
\end{align}

Second, we consider the set of $\abs{\mathcal{P}^{\textrm{SIC-POVM}}} = d^{2}$ symmetric, local, non-orthogonal, and informationally-complete Tetrad SIC-POVM measurement operators, in terms of uniformly spread out pure states $\psi_{\mu}$, $\mu \in [d^{2}]$,
\begin{align}
	\Pi_{\mu}^{\textrm{SIC-POVM}} =&~ \lambda \otimes_{i \in [n]}\psi_{\mu_{i}} \\
	\ket{\psi}_{\mu_{i}}^{\textrm{SIC-POVM}} \in&~ \{\ket{0}~,~\cos{\theta_{q}/2}\ket{0} + e^{i \mu_{i}\phi_{q}}\sin{\theta_{q}/2}\ket{1}\}_{0 < \mu_{i} < q^{2}} ~~ ~~ , ~~ ~~ \lambda = \frac{1}{q^{2n}} ~,~ \begin{array}{l}\cos{\theta_{q}/2} = \sqrt{\frac{1}{q^{2}-1}} \\ \phi_{q} = \frac{2\pi}{q^{2}-1}\end{array} \nonumber ~.
\end{align}

Third, we consider the set of $\abs{\mathcal{P}^{\textrm{NON-SIC-POVM}}} = d^{2}$ nonsymmetric, local, non-orthogonal, and informationally-complete Pauli NON-SIC-POVM measurement operators, in terms of non-uniform mixed states $\Psi_{\mu}$, $\mu \in [d^{2}]$,
\begin{align}
	\Pi_{\mu}^{\textrm{NON-SIC-POVM}} =&~ \lambda \otimes_{i \in [n]}\Psi_{\mu_{i}}^{\textrm{NON-SIC-POVM}} \\
	\Psi_{\mu_{i}}^{\textrm{NON-SIC-POVM}} \in&~ \{\ket{0}\bra{0}~,~\ket{+}\bra{+}~,~\ket{+i}\bra{+i}~,~\ket{1}\bra{1}+\ket{-}\bra{-}+\ket{-i}\bra{-i}\} ~~ \quad , \quad ~~ \lambda = \frac{1}{(q^{2}-1)^{n}}~. \nonumber
\end{align}

\subsubsection{Total distributions of measurement probabilities}
Here, we discuss an important clarifying remark, regarding the total probability over the set of operators,
\begin{align}
	P_{\mathcal{P}}(p) = \sum_{\Pi}P_{\Pi|\mathcal{P}}(\Pi)P_{\Pi}(p)
	\to
	\tilde{P}_{\mathcal{P}}(p) = \sum_{\Pi \in \mathcal{P}}P_{\Pi|\mathcal{P}}(\Pi)\tilde{P}_{\Pi}(p)
	~,
\end{align}
which is described as sampling from the joint distribution of states $\rho \sim P_{\rho}$ and operators $\Pi \sim P_{\Pi|\mathcal{P}}$. In our simulations, operators are not sampled, but deterministically iterated over all $\Pi \in \mathcal{P}$, weighted by the exact operator probability $P_{\Pi|\mathcal{P}}(\Pi)$, which we choose to be uniform $P_{\Pi|\mathcal{P}}(\Pi) = 1/\abs{\mathcal{P}}$. This partially-deterministic procedure results in solely empirical conditional distributions $\tilde{P}_{\Pi}(p)$, with a total empirical distribution of $\tilde{P}_{\mathcal{P}}(p)$. Previous works \cite{sauliere2025chaotic,gutkin2013joint} have avoided such technical considerations due to individual operators, generally projective PVM's, being symmetric, and representative of the whole set. In future works, sampling operators may be necessary for larger dimensions $d$ as exact iterations over $\tbigO{\poly{d}}$ operators become infeasible, compounding uncertainty within total empirical distributions.

\subsection{Numerical continuous variable empirical distributions}
In this work, we construct empirical distributions from numerical simulations. Here, we are considering continuous variable distributions, which given their expressions in terms of integrals as opposed to summations, are significantly more difficult to approximate than discrete variable distributions \cite{bobkov2010concentration,virkar2014power}. We must therefore make two approximations for numerical feasibility, regarding both the form of the resulting empirical distributions formed from samples, and the form of the metric used to quantify differences between distributions.

\subsubsection{Binned empirical distributions}
Regarding the empirical distributions and sampling, here we choose to perform the following binning procedure, to avoid storing exponential in system size number of samples. Given $m$ samples of continuous variables $x$, for numerical tractability when computing empirical distributions and histograms, we bin samples into $m^{\prime}$ discretized bins,
\begin{align}
	\{x_{i}\}_{i \in [m]} \to \{x^{\prime}_{i}\}_{i \in [m^{\prime}]}
	\quad : \quad x_{i} \to x^{\prime}_{i^{\prime}} ~\textnormal{if}~ x_{i}	\in [x^{\prime}_{i^{\prime}},x^{\prime}_{i^{\prime}+1}]~,
\end{align}
with bin density, bin size, and cumulative bin density,
\begin{align}
	\rho(x_{i^{\prime}}) =&~ \frac{\abs{\{x_{i} ~:~ x_{i} \in [x^{\prime}_{i^{\prime}},x^{\prime}_{i^{\prime}+1}]\}_{i \in [m]}}}{m}
	\quad , \quad
	\eta(x^{\prime}_{i}) = \frac{x^{\prime}_{i^{\prime}+1} - x^{\prime}_{i^{\prime}}}{m^{\prime}} \\
	\omega(x^{\prime}_{i^{\prime}}) =&~ \sum_{j^{\prime} \leq i^{\prime}} \rho(x_{j^{\prime}}) ~.
\end{align}
This procedure yields binned empirical distributions, in terms of indicator functions, $\indicator{x} = \left\{\begin{array}{cc} 1 & x~\textrm{is True}\\0 & x~\textrm{is False}\end{array}\right.$,
\begin{align}
	\tilde{P}^{\prime}(x) =&~ \frac{1}{m^{\prime}}\sum_{i^{\prime} \in [m^{\prime}]}\rho(x^{\prime}_{i^{\prime}})~\frac{1}{\eta(x^{\prime}_{i^{\prime}})}~\indicator{x^{\prime}_{i^{\prime}} \leq x < x^{\prime}_{i^{\prime}+1}} \\
	\tilde{F}^{\prime}(x) =&~ \sum_{i^{\prime} \in [m^{\prime}]}\omega(x^{\prime}_{i^{\prime}})~\indicator{x \geq x^{\prime}_{i^{\prime}}}~,
\end{align}
which may be interpreted in terms of the conditional distributions,
\begin{align}
	\tilde{P}(x|x_{i}) = \indicator{x=x_{i}} \quad\to\quad \tilde{P}^{\prime}(x|x^{\prime}_{i}) = \rho(x^{\prime}_{i^{\prime}})~\frac{1}{\eta(x^{\prime}_{i^{\prime}})}~\indicator{x^{\prime}_{i^{\prime}} \leq x < x^{\prime}_{i^{\prime}+1}} ~.
\end{align}
In this work's numerical studies of the measurement operator distributions $\mathcal{P}$, for efficiency given the exponentially large number of samples $m \to m \abs{\mathcal{P}}$ required in this continuous variable setting, we use binned empirical distributions. Here, we map our $m\abs{\mathcal{P}} \leq 128\abs{\mathcal{P}}$ samples to fixed $m^{\prime}=10^{4}$ binned samples of equally spaced points on a logarithmic-scale in the range $[10^{-20},1]$. Thus, $\tilde{P}(x) \to \tilde{P}^{\prime}(x)$, and $\tilde{F}(x) \to \tilde{F}^{\prime}(x)$ are implicitly replaced in any expressions involving the empirical distributions. Such binning introduces bias into the empirical distributions, possibly masks sample complexity effects with $m$, and ultimately complicates sample complexity analysis \cite{bobkov2010concentration}. However, given smooth enough distributions, an appropriately chosen logarithmic-scale for the binning, which is shown to affect bias less \cite{virkar2014power}, and given large enough $m^{\prime} \gg m$, such binning should not significantly affect any interpretations. \\

In \cref{fig:plot_probability_sic_povm,fig:plot_probability_nonsic_povm} we show the resulting respective Tetrad SIC-POVM and Pauli NON-SIC-POVM binned empirical distribution histograms for various system parameters. Also plotted is the fit effective analytical models for comparison to the empirical distributions, for the SIC-POVM distributions.

\clearpage
\newpage

\twocolumngrid

\begin{figure}[ht]
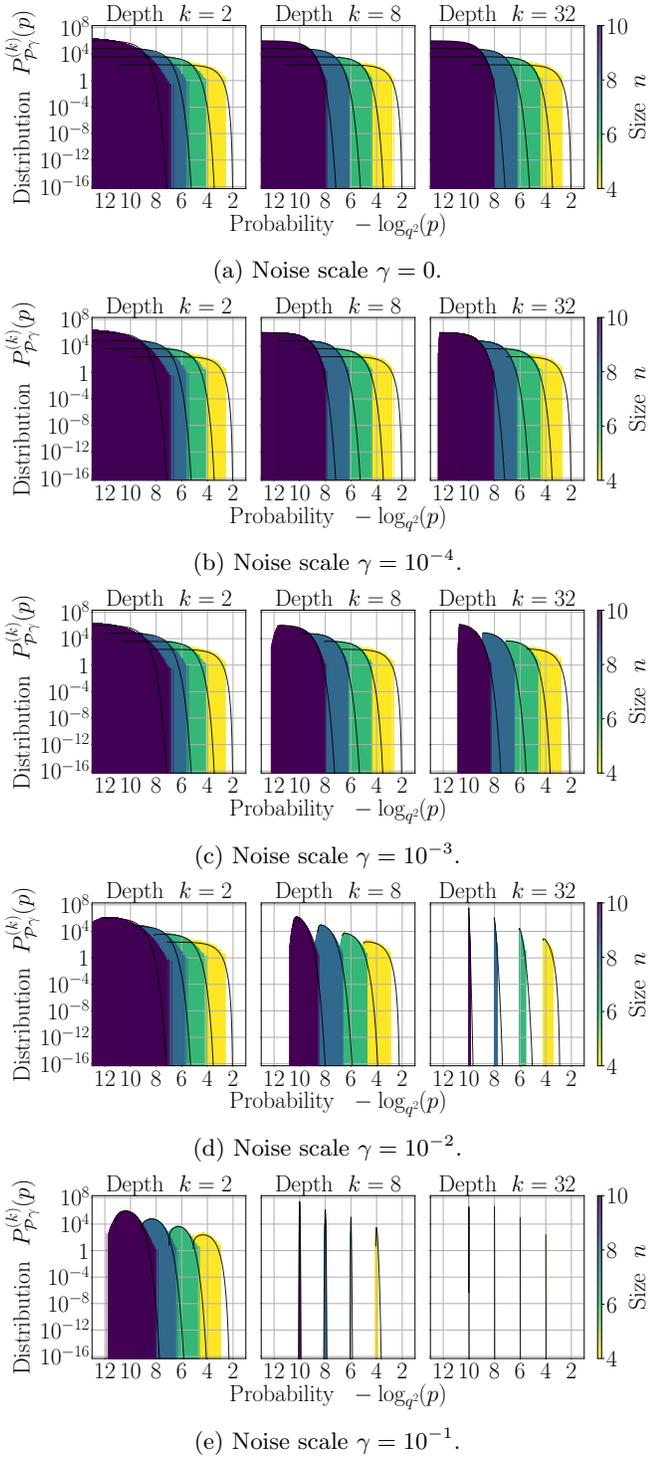

	\centering
	\begin{subfigure}[t]{\columnwidth}
		\centering
		\includegraphics[width=1\columnwidth]{plot.sample.array.M.noise.parameters.N.tetrad.0.pdf}
		\subcaption{Noise scale $\gamma = 0$.}
		\label{fig:plot_probability_sic_povm_0}
	\end{subfigure}
	\hfill
	\begin{subfigure}[t]{\columnwidth}
		\centering
		\includegraphics[width=1\columnwidth]{plot.sample.array.M.noise.parameters.N.tetrad.4.pdf}
		\subcaption{Noise scale $\gamma = 10^{-4}$.}
		\label{fig:plot_probability_sic_povm_4}
	\end{subfigure}
	\hfill
	\begin{subfigure}[t]{\columnwidth}
		\centering
		\includegraphics[width=1\columnwidth]{plot.sample.array.M.noise.parameters.N.tetrad.3.pdf}
		\subcaption{Noise scale $\gamma = 10^{-3}$.}
		\label{fig:plot_probability_sic_povm_3}
	\end{subfigure}
	\hfill
	\begin{subfigure}[t]{\columnwidth}
		\centering
		\includegraphics[width=1\columnwidth]{plot.sample.array.M.noise.parameters.N.tetrad.2.pdf}
		\subcaption{Noise scale $\gamma = 10^{-2}$.}
		\label{fig:plot_probability_sic_povm_2}
	\end{subfigure}
	\hfill
	\begin{subfigure}[t]{\columnwidth}
		\centering
		\includegraphics[width=1\columnwidth]{plot.sample.array.M.noise.parameters.N.tetrad.1.pdf}
		\subcaption{Noise scale $\gamma = 10^{-1}$.}
		\label{fig:plot_probability_sic_povm_1}
	\end{subfigure}
	\captionsetup{justification=raggedright}
	\caption{Empirical SIC-POVM probability histograms $\tilde{P}^{(k)}_{\mathcal{P}\gamma}(p)$ for depths $k$ (rows), noise scales $\gamma$ (a-e), system sizes $n$ (colored histograms), and $m=128$ samples, for Haar random brickwork and local depolarization circuits. Solid lines indicate the fit effective analytical distribution $P_{\mathcal{P}\gamma}^{(k)}$ for each system size. The effective analytical models generally capture the distributions' peaks, particularly as depth or noise scales increase and the distribution peaks sharpen and tails decrease.}
	\label{fig:plot_probability_sic_povm}
\end{figure}
\begin{figure}[htp]
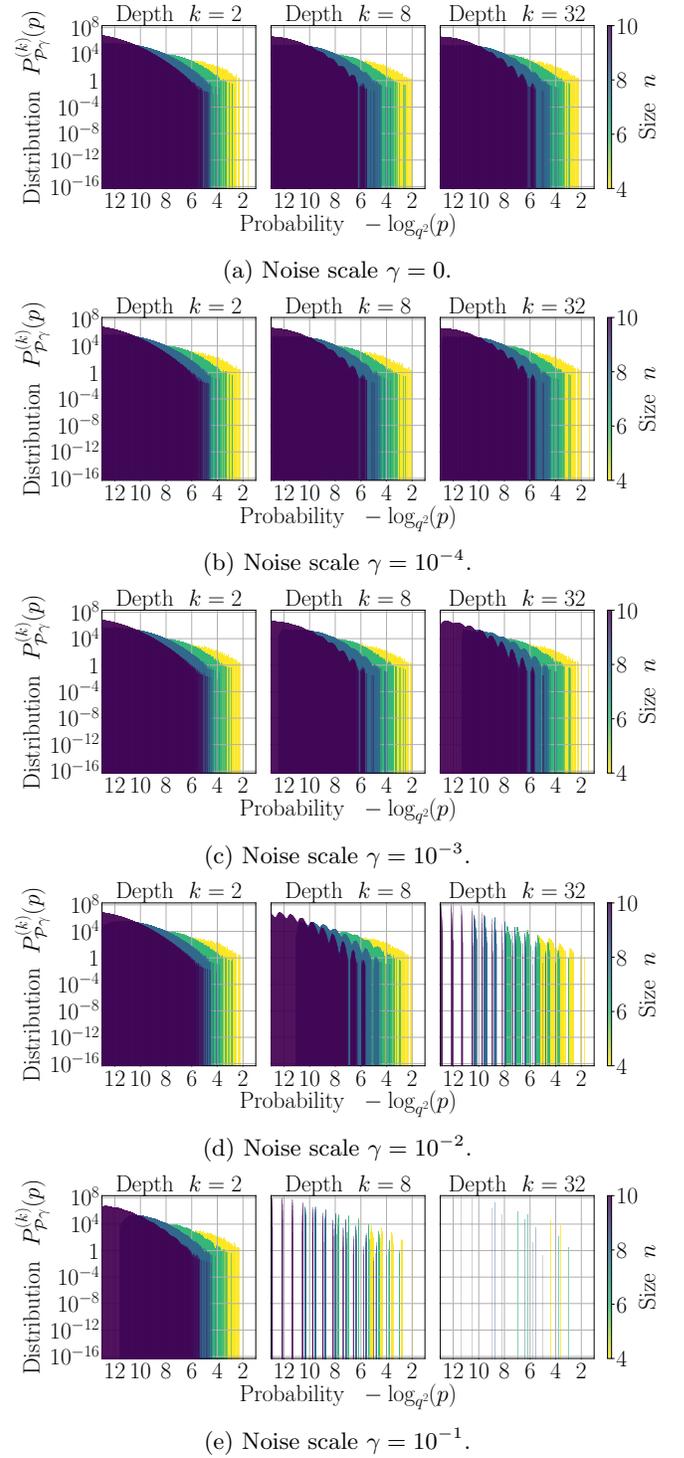

	\centering
	\begin{subfigure}[t]{\columnwidth}
		\centering
		\includegraphics[width=1\columnwidth]{plot.sample.array.M.noise.parameters.N.pauli.0.pdf}
		\subcaption{Noise scale $\gamma = 0$.}
		\label{fig:plot_probability_nonsic_povm_0}
	\end{subfigure}
	\hfill
	\begin{subfigure}[t]{\columnwidth}
		\centering
		\includegraphics[width=1\columnwidth]{plot.sample.array.M.noise.parameters.N.pauli.4.pdf}
		\subcaption{Noise scale $\gamma = 10^{-4}$.}
		\label{fig:plot_probability_nonsic_povm_4}
	\end{subfigure}
	\hfill
	\begin{subfigure}[t]{\columnwidth}
		\centering
		\includegraphics[width=1\columnwidth]{plot.sample.array.M.noise.parameters.N.pauli.3.pdf}
		\subcaption{Noise scale $\gamma = 10^{-3}$.}
		\label{fig:plot_probability_nonsic_povm_3}
	\end{subfigure}
	\hfill
	\begin{subfigure}[t]{\columnwidth}
		\centering
		\includegraphics[width=1\columnwidth]{plot.sample.array.M.noise.parameters.N.pauli.2.pdf}
		\subcaption{Noise scale $\gamma = 10^{-2}$.}
		\label{fig:plot_probability_nonsic_povm_2}
	\end{subfigure}
	\hfill
	\begin{subfigure}[t]{\columnwidth}
		\centering
		\includegraphics[width=1\columnwidth]{plot.sample.array.M.noise.parameters.N.pauli.1.pdf}
		\subcaption{Noise scale $\gamma = 10^{-1}$.}
		\label{fig:plot_probability_nonsic_povm_1}
	\end{subfigure}
	\captionsetup{justification=raggedright}
	\caption{Empirical NON-SIC-POVM probability histograms $\tilde{P}^{(k)}_{\mathcal{P}\gamma}(p)$ for depths $k$ (rows), noise scales $\gamma$ (a-e), system sizes $n$ (colored histograms), and $m=128$ samples, for Haar random brickwork and local depolarization circuits.
	As depth and noise scales increase, the empirical distributions develop many sharpened peaks, due to being a sum of distributions over all nonsymmetric measurement operators, each contributing shifted and peaked conditional distributions to the total distribution.}
	\label{fig:plot_probability_nonsic_povm}
\end{figure}

\clearpage
\newpage

\onecolumngrid

\subsubsection{Empirical Kolmogorov–Smirnov metrics}
Regarding metrics to quantify differences between distributions, here we use an upper bound on the Kolmogorov–Smirnov metric to avoid maximizations over continuous variable domains. Given the monotonicity property of cumulative distributions, then $F(x_{i}) \leq F(x_{i+1}) ~\forall~ x_{i} \leq x_{i+1}$, and similarly given the piecewise-constant property of empirical distributions, then $\tilde{F}(x) = \tilde{F}(x_{i}) ~\forall~ x \in [x_{i},x_{i+1}]$. From these properties, given $m$ samples $\{x_{i}\}_{i \in [m]}$, and $x \in [x_{i},x_{i+1}]$, then the difference of empirical and analytical distributions is,
\begin{align}
	\abs{\tilde{F}(x) - F(x)}
	=&~ \abs{[\tilde{F}(x) - F(x_{i}] - [F(x)-F(x_{i})]} \\
	\leq&~ \abs{\tilde{F}(x) - F(x_{i})} + \abs{F(x)-F(x_{i})} \\
	\leq&~ \abs{\tilde{F}(x_{i}) - F(x_{i})} + \abs{F(x_{i+1})-F(x_{i})}~.
\end{align}
Optimizations over the entire domain $x$ can be replaced by the maximization over the $m$ samples, yielding what we refer to as the empirical Kolmogorov–Smirnov metric $\tilde{\mathcal{L}}$,
\begin{align}
	\mathcal{L} = \max_{x} \abs{\tilde{F}(x) - F(x)} \leq \max_{i \in [m]}~ \abs{\tilde{F}(x_{i}) - F(x_{i})} ~+~ \abs{F(x_{i+1})-F(x_{i})} \equiv \tilde{\mathcal{L}}~.
\end{align}
Depending on the smoothness of the distributions, such an upper bound, with the additional difference term $\abs{F(x_{i+1})-F(x_{i})}$, may not necessarily be tight. However, general convergence trends should be demonstrated for sufficiently large $m$ samples and carefully binned $m^{\prime}$ bins as $\lim_{x_{i+1} \to x_{i}} F(x_{i+1}) \to F(x_{i})$ tightens.

Returning to our measurement probability distributions, to assess the similarity of empirical distributions $\tilde{F}^{(k)}_{\mathcal{P}\gamma}$ of measurement probabilities $p$ to analytical distributions ${F}^{(k)}_{\mathcal{P}\gamma}$, we compute the empirical Kolmogorov–Smirnov metric,
\begin{align}
	\mathcal{L}_{\mathcal{P}\gamma}^{(k)}
	\leq&~ \tilde{\mathcal{L}}^{(k)}_{\mathcal{P}\gamma} = \max_{i \in [m]} ~\abs{\tilde{F}^{(k)}_{\mathcal{P}\gamma}(p_{i})-{F}^{(k)}_{\mathcal{P}\gamma}(p_{i})}~+~\abs{{F}^{(k)}_{\mathcal{P}\gamma}(p_{i+1})-{F}^{(k)}_{\mathcal{P}\gamma}(p_{i})} \\
	\approx&~ \tilde{\mathcal{L}}^{(k)\prime}_{\mathcal{P}\gamma} = \max_{i \in [m^{\prime}]} ~\abs{\tilde{F}^{(k)}_{\mathcal{P}\gamma}(p_{i}^{\prime})-{F}^{(k)}_{\mathcal{P}\gamma}(p_{i}^{\prime})}~+~\abs{{F}^{(k)}_{\mathcal{P}\gamma}(p_{i+1}^{\prime})-{F}^{(k)}_{\mathcal{P}\gamma}(p_{i}^{\prime})} ~,
\end{align}
which we further approximate using our binned empirical distribution $\tilde{\mathcal{L}}^{(k)\prime}_{\mathcal{P}\gamma}$ using our $m^{\prime}$ binned samples.

\subsubsection{Sample complexity of empirical distributions}
It remains to be seen how the number of samples $m$ affects the behavior of these metrics, in particular given a fixed-size $m^{\prime}$ binning procedure that is independent of $m$. In \cref{fig:plot_metric_sic_povm_32_64_128}, we plot the empirical, binned Kolmogorov–Smirnov metric for $m \in \{32,64,128\}$ number of samples. Such metrics appear very consistent across number of samples, with slightly more stable, and more convergent behavior for the noiseless simulations at larger sample sizes.

The consistency of the empirical Kolmogorov–Smirnov metrics across number of samples is potentially due to masking of sampling effects. First, the exact deterministic operator sampling, which multiplies the number of samples to be $m \to m\abs{\mathcal{P}}$, with $\abs{\mathcal{P}} \gg m$, may mask any $m$-dependencies. Second, the binning procedure on top of any sampling may predominantly mask any sampling procedures, particularly in parameter regimes where there are highly peaked distribution. Even with fine discretization into $m^{\prime} \gg m$ uniform logarithmically spaced bins, non-uniform bins around the peaks and tails may be more appropriate to better understand sample complexities.

Given these biases, care must be taken in interpreting Chebyshev's inequality of the empirical distribution $\tilde{F}$ differing from its mean, as this mean is not necessarily $F$, but whichever distribution the empirical distribution truly converges towards. However, given the variance of the empirical distribution itself is always bounded by a constant independent of the system when $F = 1/2$, therefore the system-independent sample complexity should still hold,
\begin{align}
	m \geq \left(\frac{1}{2\delta\epsilon}\right)^{2}~,
\end{align}
whose system independence could also partially explain the consistency of our metric across samples $m$ and sizes $n$.~\\

\begin{figure}[thp]
	\centering
	\begin{subfigure}[t]{0.32\columnwidth}
		\centering
		\includegraphics[width=\textwidth]{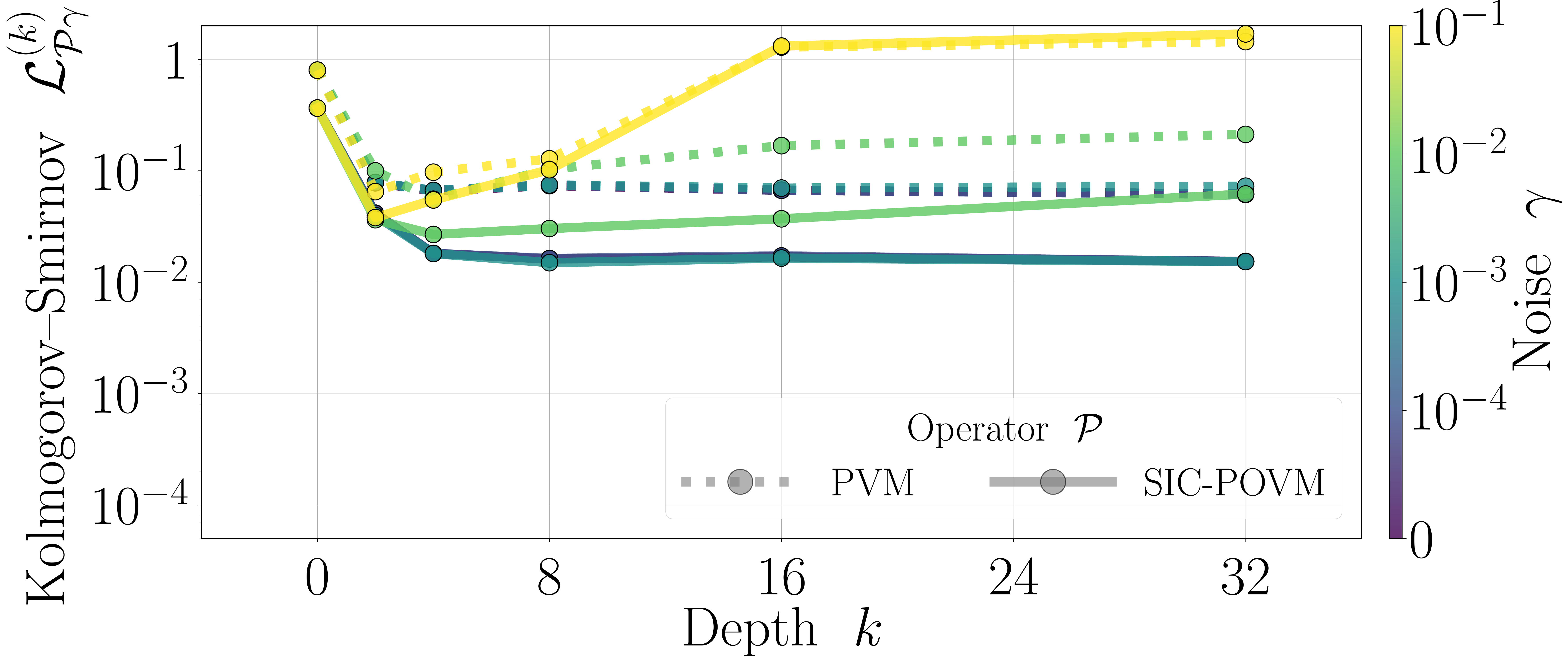}
		\subcaption{Size $n=4$.}
		\label{fig:plot_metric_sic_povm_4_32}
	\end{subfigure}
	\begin{subfigure}[t]{0.32\columnwidth}
		\centering
		\includegraphics[width=\textwidth]{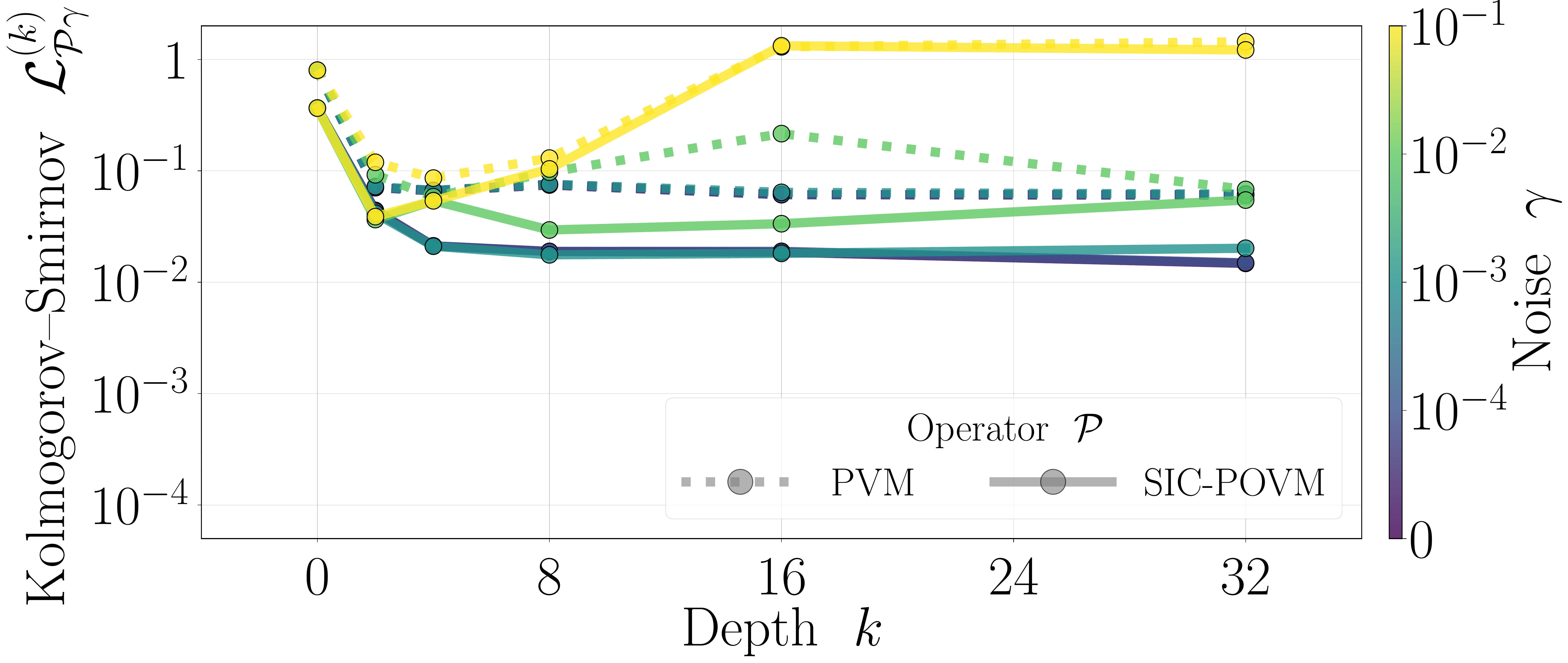}
		\subcaption{Size $n=4$.}
		\label{fig:plot_metric_sic_povm_4_64}
	\end{subfigure}
	\begin{subfigure}[t]{0.32\columnwidth}
		\centering
		\includegraphics[width=\textwidth]{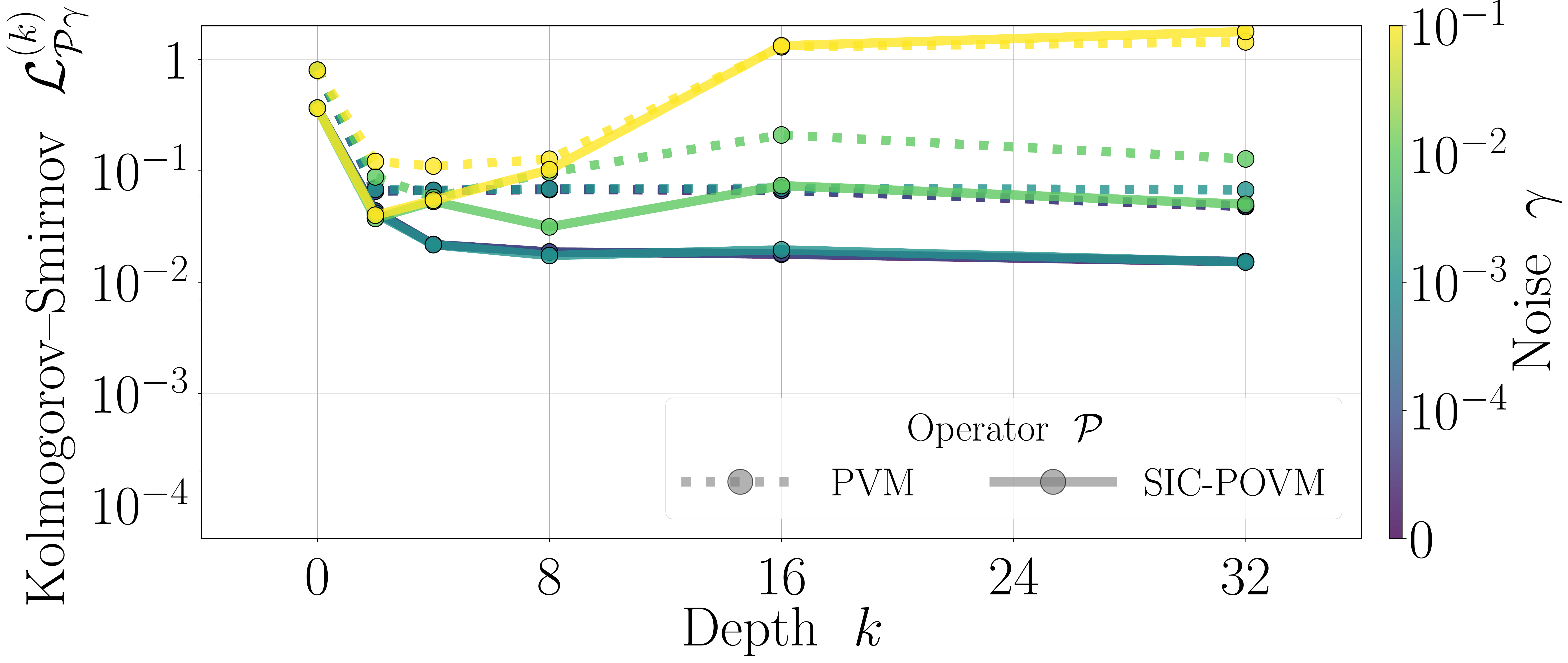}
		\subcaption{Size $n=4$.}
		\label{fig:plot_metric_sic_povm_4_128}
	\end{subfigure}
	\begin{subfigure}[t]{0.32\columnwidth}
		\centering
		\includegraphics[width=\textwidth]{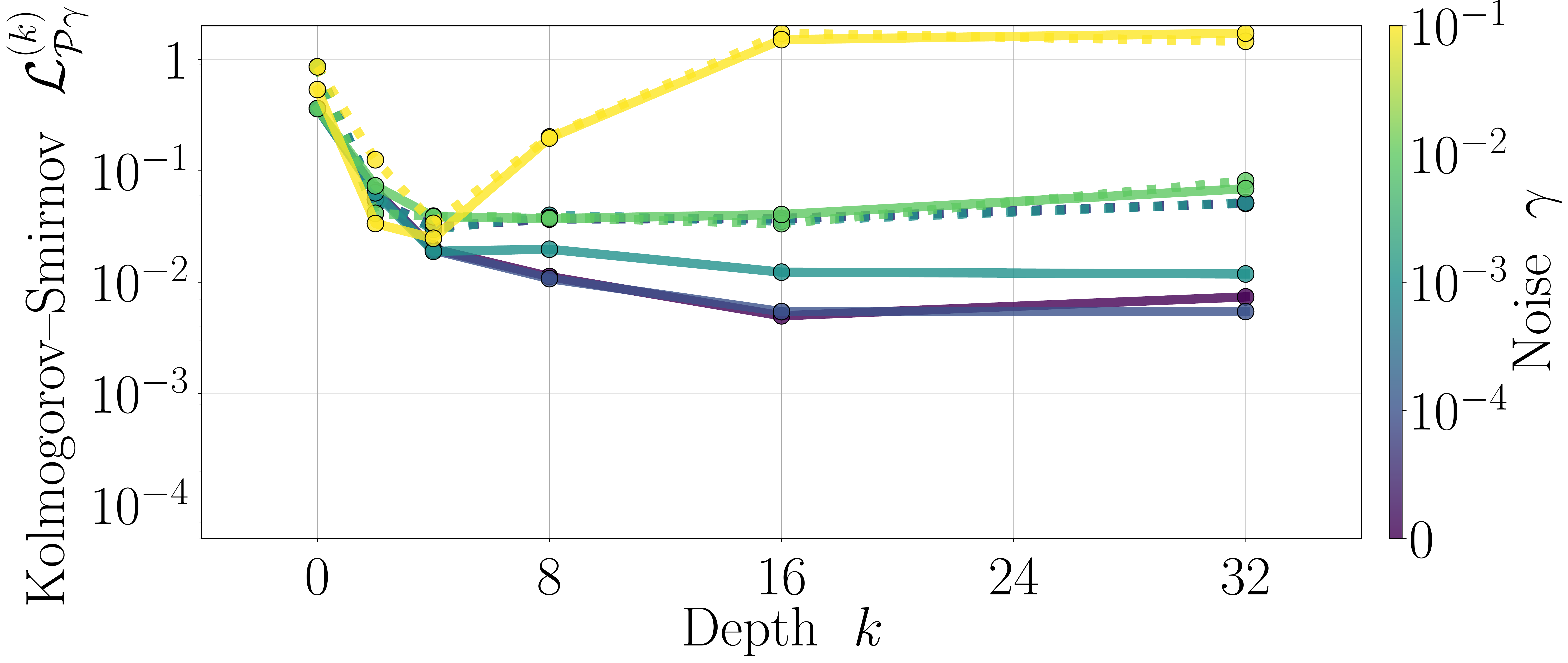}
		\subcaption{Size $n=6$.}
		\label{fig:plot_metric_sic_povm_6_32}
	\end{subfigure}
	\begin{subfigure}[t]{0.32\columnwidth}
		\centering
		\includegraphics[width=\textwidth]{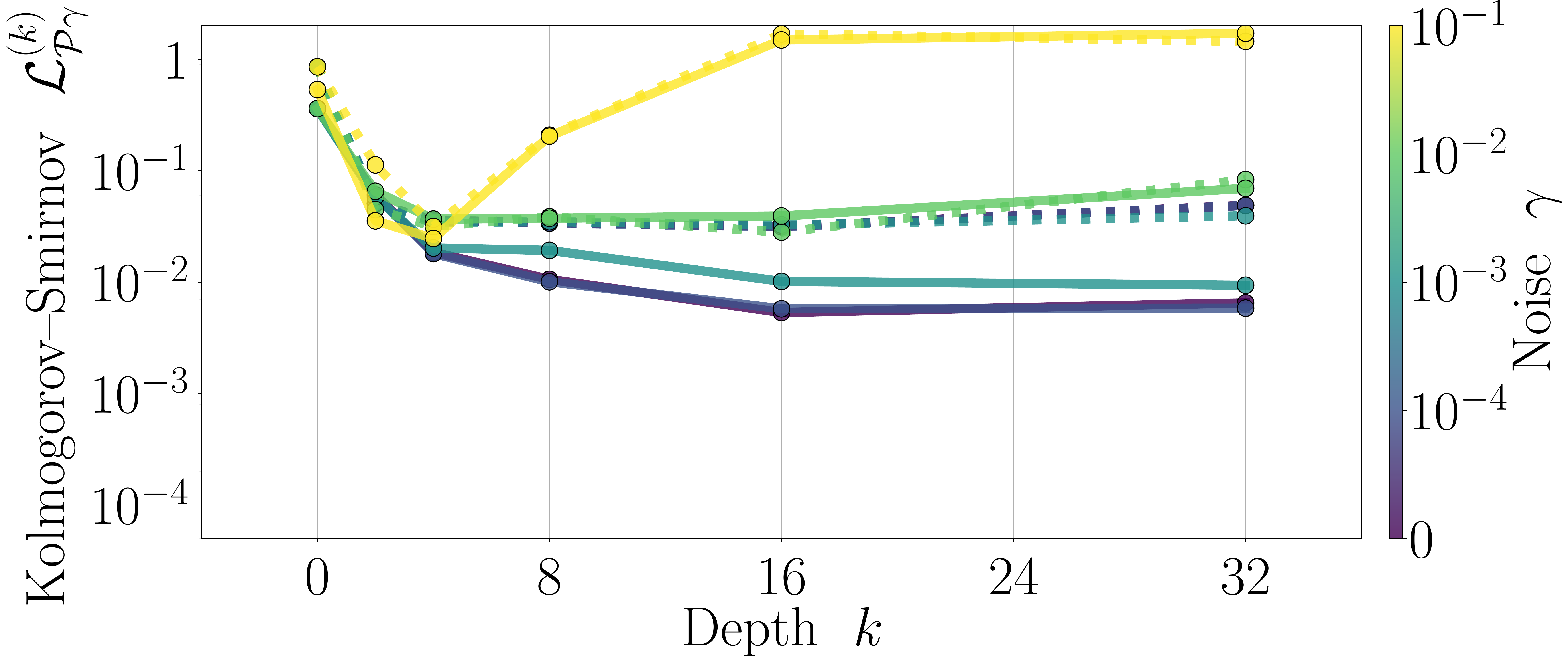}
		\subcaption{Size $n=6$.}
		\label{fig:plot_metric_sic_povm_6_64}
	\end{subfigure}
	\begin{subfigure}[t]{0.32\columnwidth}
		\centering
		\includegraphics[width=\textwidth]{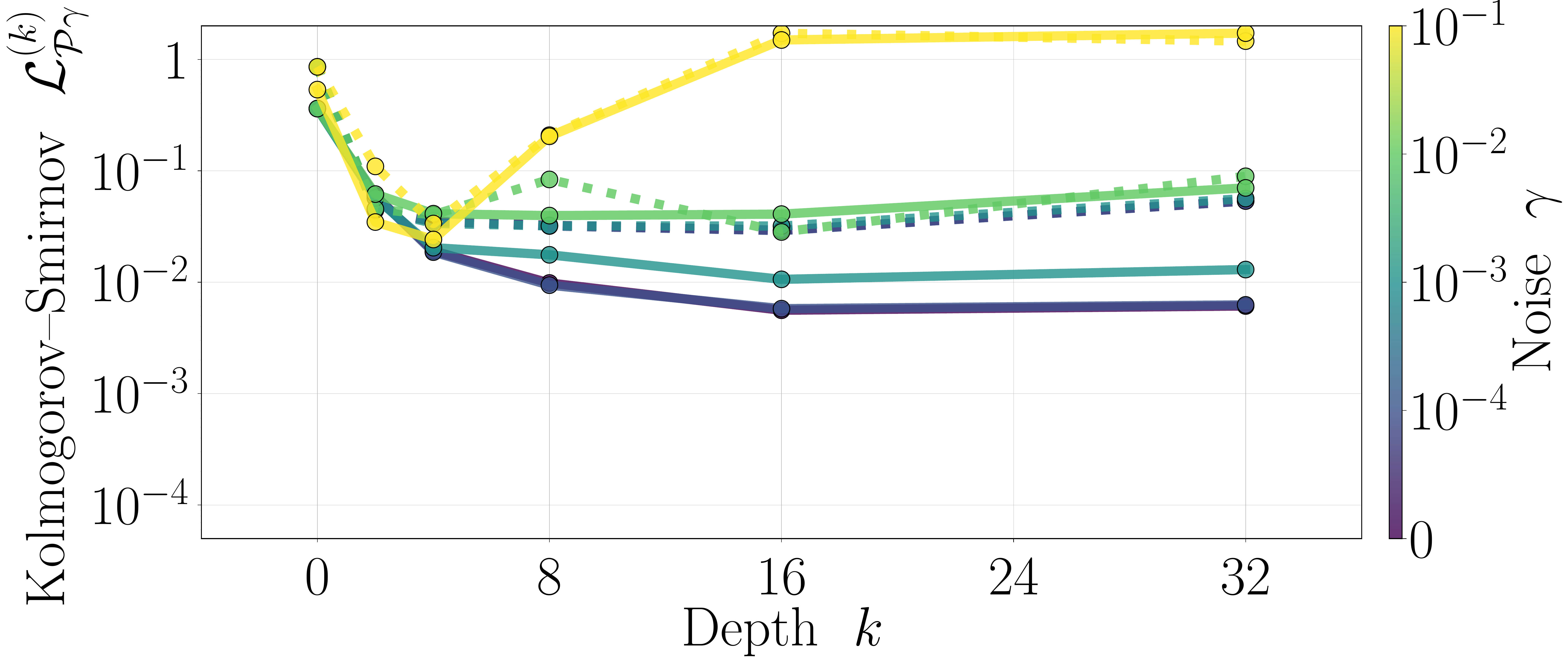}
		\subcaption{Size $n=6$.}
		\label{fig:plot_metric_sic_povm_6_128}
	\end{subfigure}
	\hfill
	\begin{subfigure}[t]{0.32\columnwidth}
		\centering
		\includegraphics[width=\textwidth]{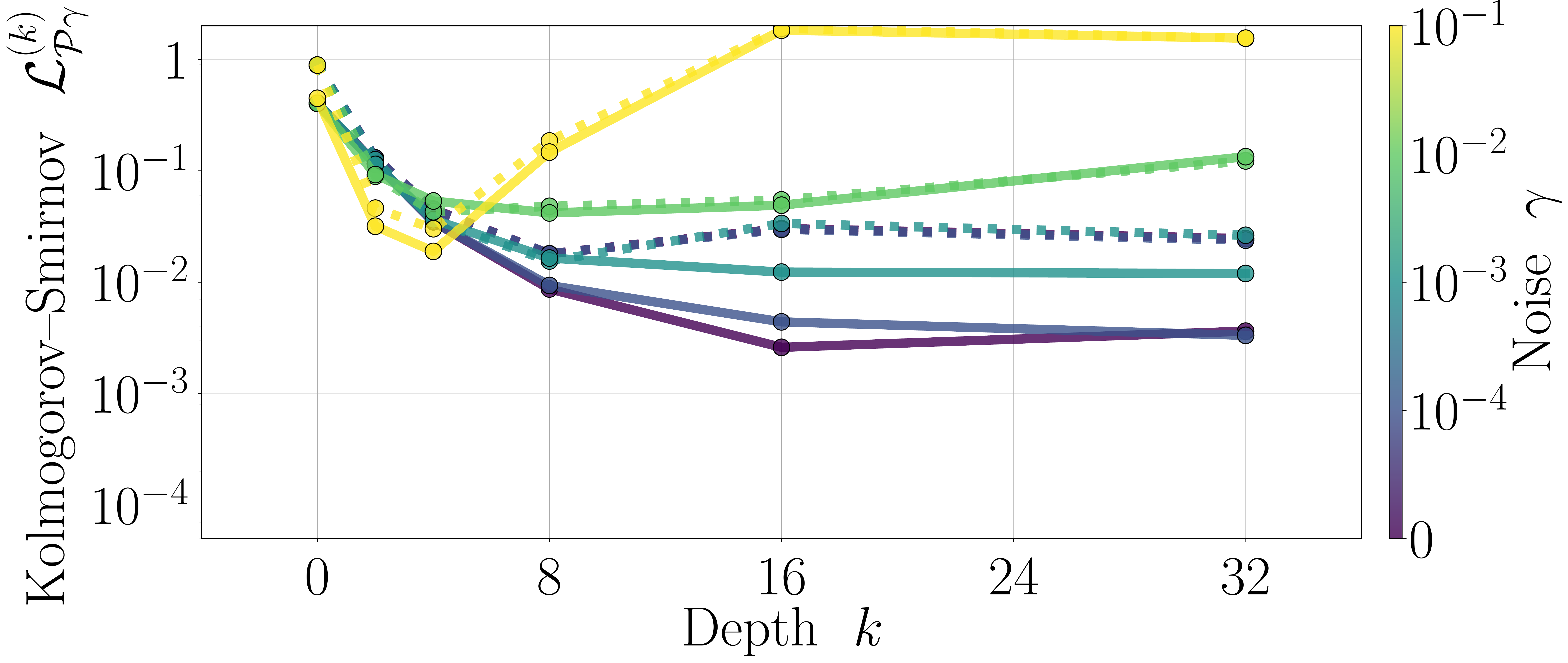}
		\subcaption{Size $n=8$.}
		\label{fig:plot_metric_sic_povm_8_32}
	\end{subfigure}
	\begin{subfigure}[t]{0.32\columnwidth}
		\centering
		\includegraphics[width=\textwidth]{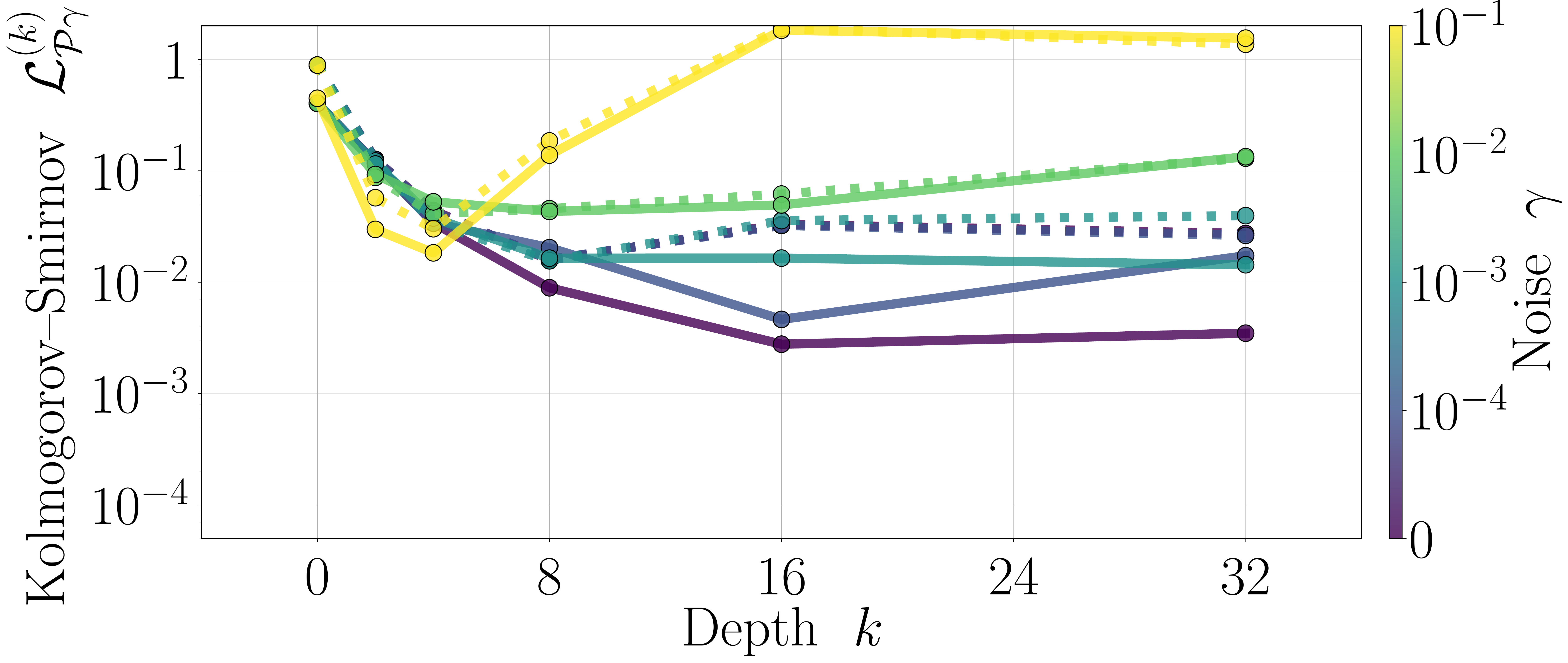}
		\subcaption{Size $n=8$.}
		\label{fig:plot_metric_sic_povm_8_64}
	\end{subfigure}
	\begin{subfigure}[t]{0.32\columnwidth}
		\centering
		\includegraphics[width=\textwidth]{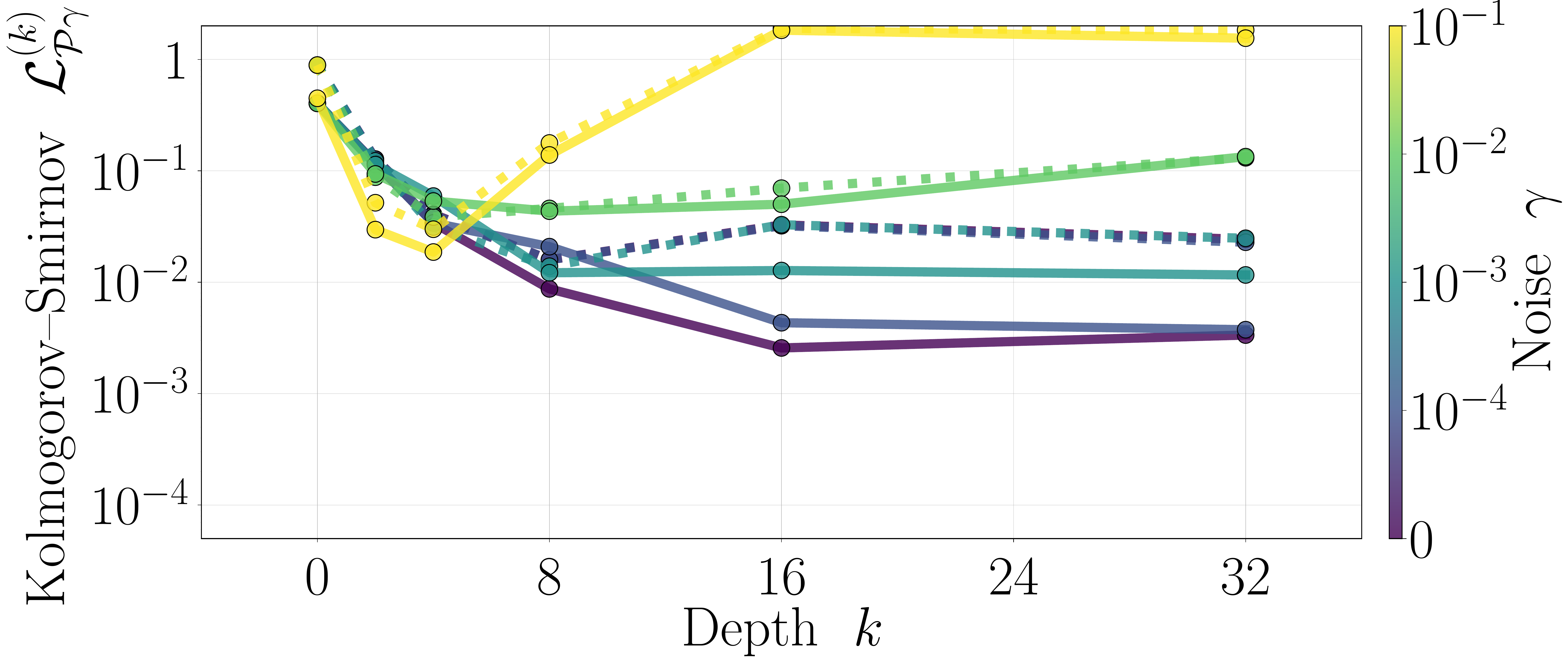}
		\subcaption{Size $n=8$.}
		\label{fig:plot_metric_sic_povm_8_128}
	\end{subfigure}
	\hfill
	\begin{subfigure}[t]{0.32\columnwidth}
		\centering
		\includegraphics[width=\textwidth]{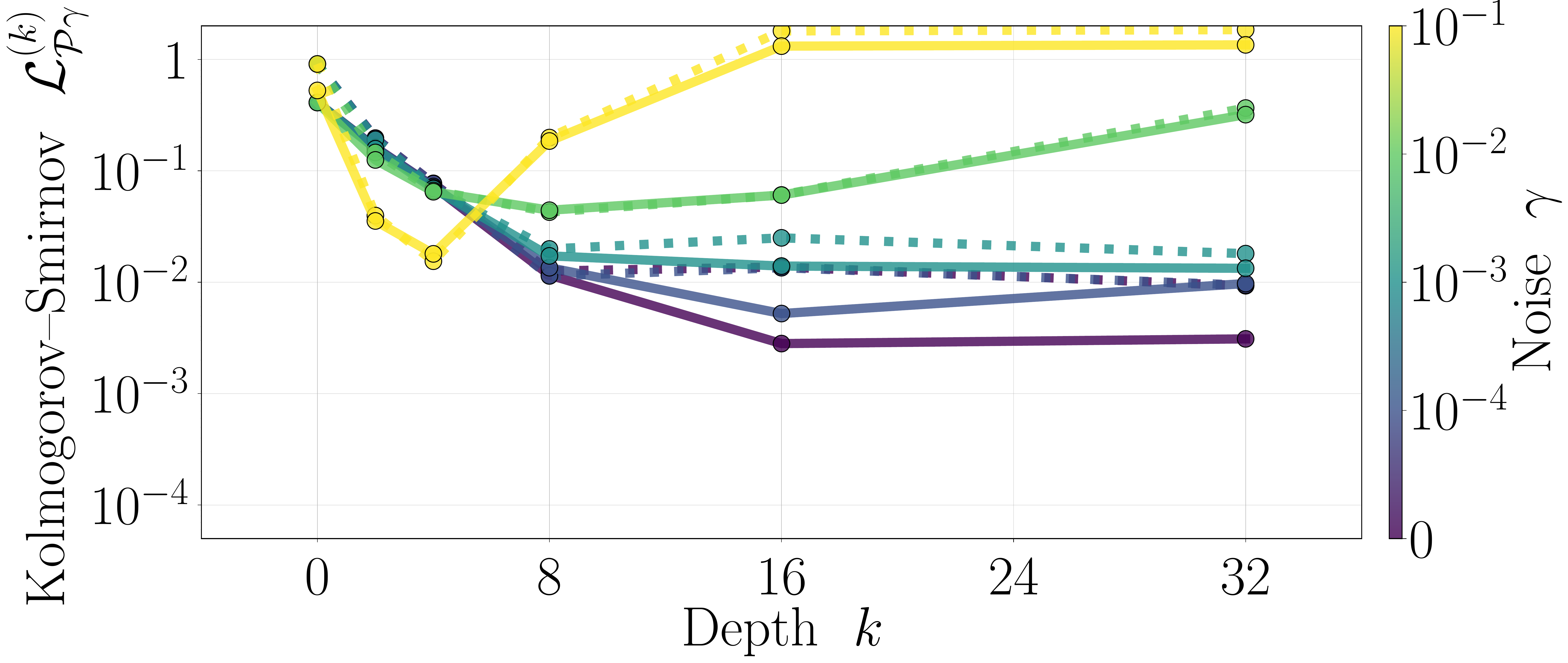}
		\subcaption[Size $n=10$,$m=32$ samples.]{Size $n=10$\protect\linebreak\protect\linebreak$m=32$ samples.}
		\label{fig:plot_metric_sic_povm_10_32}
	\end{subfigure}
	\begin{subfigure}[t]{0.32\columnwidth}
		\centering
		\includegraphics[width=\textwidth]{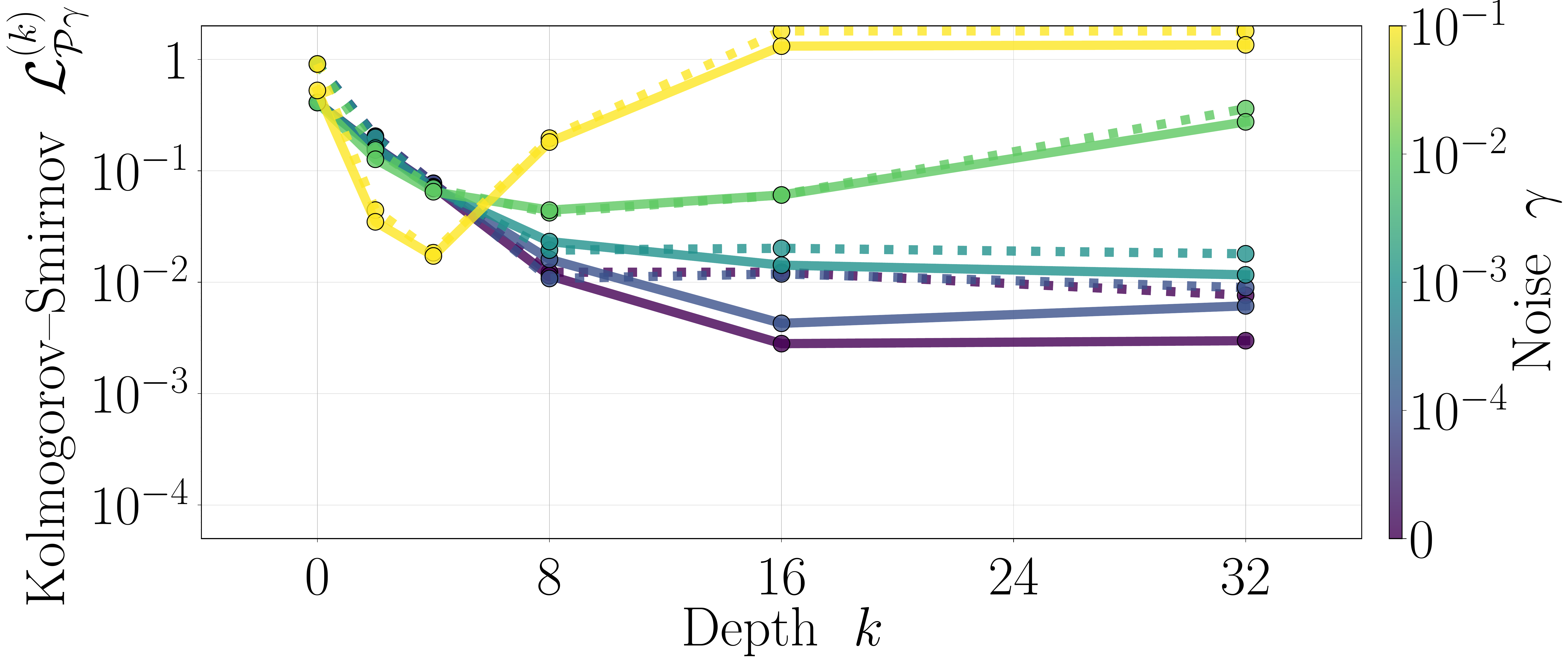}
		\subcaption[Size $n=10$,$m=64$ samples.]{Size $n=10$\protect\linebreak\protect\linebreak$m=64$ samples.}
		\label{fig:plot_metric_sic_povm_10_64}
	\end{subfigure}
	\begin{subfigure}[t]{0.32\columnwidth}
		\centering
		\includegraphics[width=\textwidth]{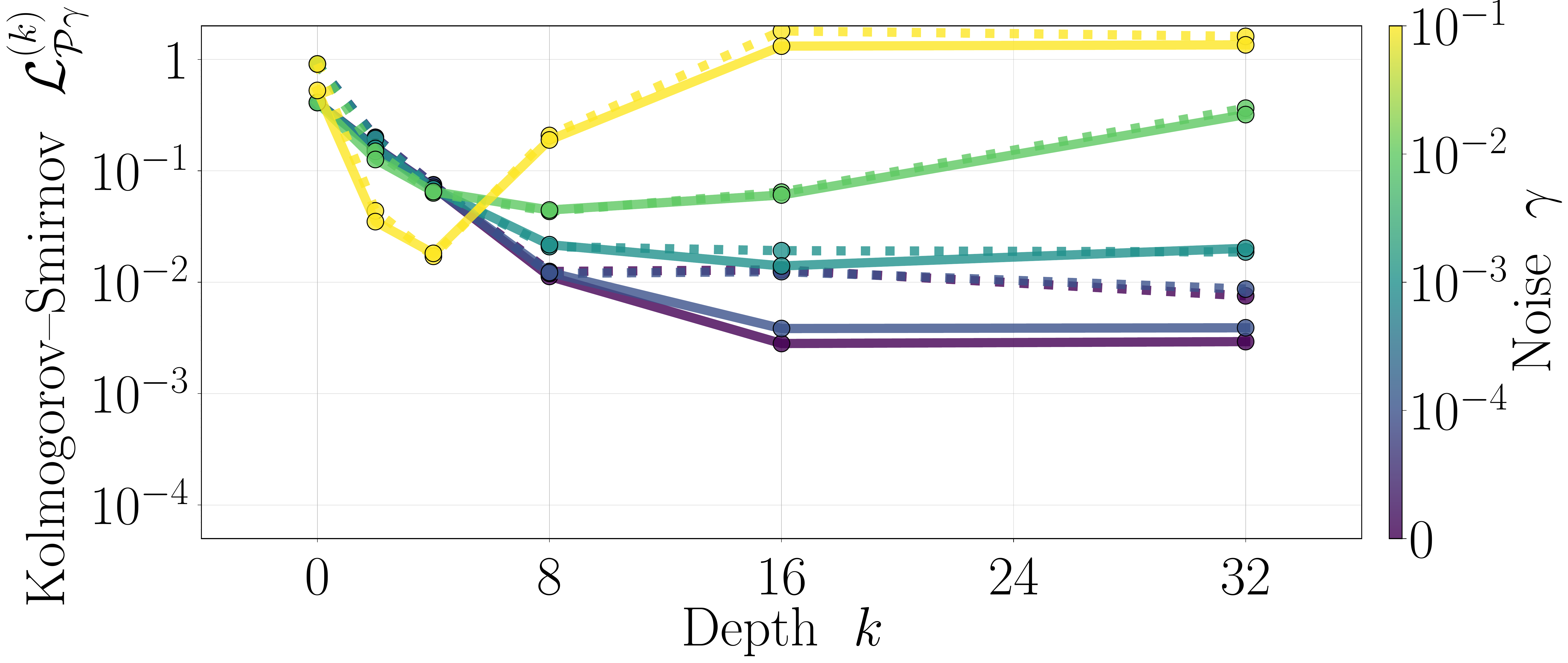}
		\subcaption[Size $n=10$,$m=128$ samples]{Size $n=10$\protect\linebreak\protect\linebreak$m=128$ samples.}
		\label{fig:plot_metric_sic_povm_10_128}
	\end{subfigure}
	\captionsetup{justification=raggedright}
	\caption{Sampling-dependence of empirical Kolmogorov–Smirnov metric $\mathcal{L}_{\mathcal{P}\gamma}$ of upper-bounded maximum difference between empirical and analytical cumulative distributions for SIC-POVM and PVM distributions $\tilde{F}^{(k)}_{\mathcal{P}\gamma}(p)$, as a function of depth $k$, for noise scales $\gamma$ (colors), system sizes $n$ (rows), and $m$ (columns) samples, for Haar random brickwork and local depolarization circuits. As depth increases, the metrics converge towards zero, before plateauing due to inherent biases in the binning procedures, and in the effective models for $F^{(k)}_{\mathcal{P}\gamma}(p)$.}
	\label{fig:plot_metric_sic_povm_32_64_128}
\end{figure}

\subsection{Effective analytical models}
In this work, given our analytical models of noiseless distributions, we propose an effective analytical model for noisy distributions, and interpret the behavior of the model's parameters as a function of system parameters.

\subsubsection{Selection of effective analytical models}
Given our simulated systems, namely brickwork quantum circuits with interspersed local depolarizing noise, we must propose an appropriate effective analytical model for the distribution of its resulting expectation values. In fact, such circuits have been studied analytically recently \cite{sauliere2026universality,sauliere2025chaotic}, where expressions for moments in asymptotic limits are derived, and distributions are constructed from fitting procedures and truncated series expansions. Here, we desire immediately interpretable closed-form expressions for distributions. We thus will take inspiration from recent results that noisy random quantum circuit probabilities converge to the globally depolarized uniform distribution, within shallow circuit depths scaling logarithmically with system size \cite{dalzell2021random,deshpande2022tight}.

In particular, although developing analytical local noise and local Haar random state models appears out of the scope of this work, we can develop effective global noise and global Haar random state models, which over certain noise scales and circuit depths, approximates the noisy behaviors well. Here, we use our previous insight that isotropic depolarizing-like quantum channels have simple and intuitive shifted and scaled distributions,
\begin{align}
	\Pi \to (1-\tilde{\gamma})\Pi + \tilde{\gamma} \frac{\trace{\Pi}}{d} I
	\quad\quad : \quad\quad
	P_{\Pi}(p) \to&~ \frac{1}{1-\tilde{\gamma}}P_{\Pi}\left(\frac{p-\tilde{\gamma}\trace{\Pi}/d}{1-\tilde{\gamma}}\right) ~\equiv~ {P}_{\Pi\gamma}^{(k)}(p) ~\approx~ \tilde{P}_{\Pi\gamma}^{(k)}(p)~.
\end{align}

\newpage
Here, we define the arbitrary noise scale $\tilde{\gamma} = \tilde{\gamma}(k,\gamma,n,m)$ and environment dimension $\tilde{s} = \tilde{s}(k,\gamma,n,m)$ as variable effective model parameters, to be fit using empirical data samples $\tilde{P}_{\Pi\gamma}^{(k)}(p_{i})$, as functions of system parameters $k,\gamma,n,m$.

\subsubsection{Optimization of effective analytical models}
To find the optimal effect model parameters $\tilde{\gamma},\tilde{s}$, we minimize the constrained normalized mean-squared-error between the (binned) empirical distribution and the model, evaluated at the $m \to m^{\prime}$ samples of empirical data,
\begin{align}
	\tilde{\gamma},\tilde{s} =&~ ~\displaystyle\argmin_{\tilde{\gamma},\tilde{s}}~~ \frac{1}{m^{\prime}}\frac{\sum_{i \in [m^{\prime}]}\abs{{P}_{\mathcal{P}\gamma}^{(k)}(p^{\prime}_{i}|\tilde{\gamma},\tilde{s}) - \tilde{P}_{\mathcal{P}\gamma}^{(k)}(p^{\prime}_{i})}^{2}}{\sum_{i \in [m^{\prime}]}\abs{\tilde{P}_{\mathcal{P}\gamma}^{(k)}(p^{\prime}_{i})}^{2}}
	\quad \quad : \quad \quad \begin{array}{l} 0 \leq \tilde{\gamma} \leq 1 \\ 1 \leq \tilde{s} \leq d^{2} \end{array} ~.
\end{align}
We numerically perform such optimizations using the \emph{scipy.optimize.minimize} optimizer \cite{virtanen2020scipy}, with the following options: 	\emph{method}: $None$, \emph{bounds}: $\tilde{\gamma} \in [0,0.99999999]$, $\tilde{s} \in [1,\infty]$, \emph{tol}: $10^{-16}$, \emph{ftol}: $10^{-16}$, \emph{gtol}: $10^{-16}$, \emph{eps}: $10^{-8}$, \emph{maxiter}: $1000$, \emph{maxls}: $64$, and did not conduct a full hyper-parameter search. The optimizer reported successful, converged optimizations according to these criteria, however it was not verified whether global optima are reached.

Given such optimizations, under the assumption of convergence, we can interpret how the effective model parameters vary with system parameters. In the noiseless limit, we recover the expected consistent behavior, indicating the optimization is within a physically consistent local minimum,
\begin{align}
	\lim_{\gamma \to 0}~~\tilde{\gamma} \to 0 ~,~ \tilde{s} \to 1~.
\end{align}

\begin{figure}[hp]
	\centering
	\begin{subfigure}[t]{0.49\columnwidth}
		\centering
		\includegraphics[width=1\textwidth]{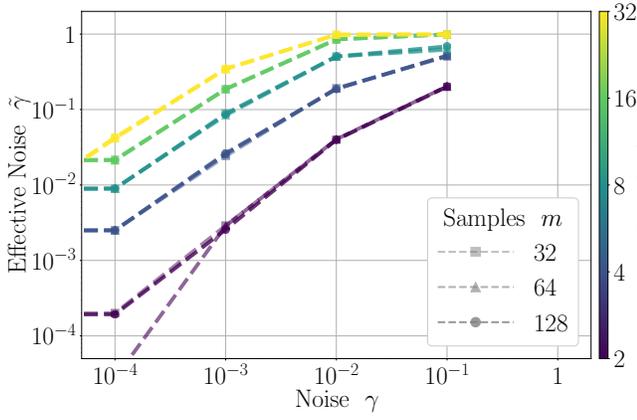}
		\subcaption{Effective noise $\tilde{\gamma}$ as a function of noise $\gamma$.}
		\label{fig:plot_parameters_sic_povm_noise_noise}
	\end{subfigure}
	\begin{subfigure}[t]{0.49\columnwidth}
		\centering
		\includegraphics[width=1\textwidth]{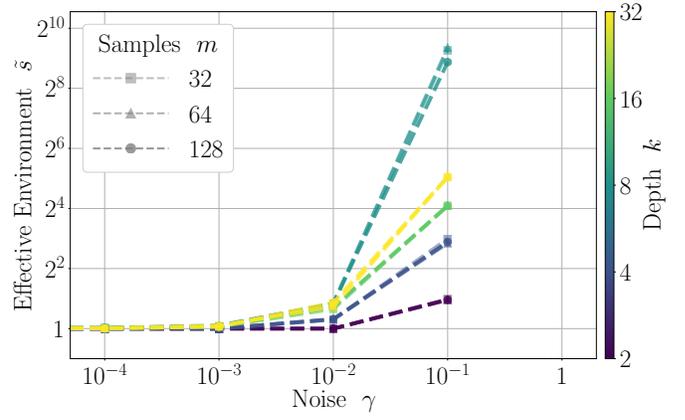}
		\subcaption{Effective environment $\tilde{s}$ as a function of noise $\gamma$.}
		\label{fig:plot_parameters_sic_povm_noise_env}
	\end{subfigure}
	\hfill
	\begin{subfigure}[t]{0.49\columnwidth}
		\centering
		\includegraphics[width=1.01\textwidth]{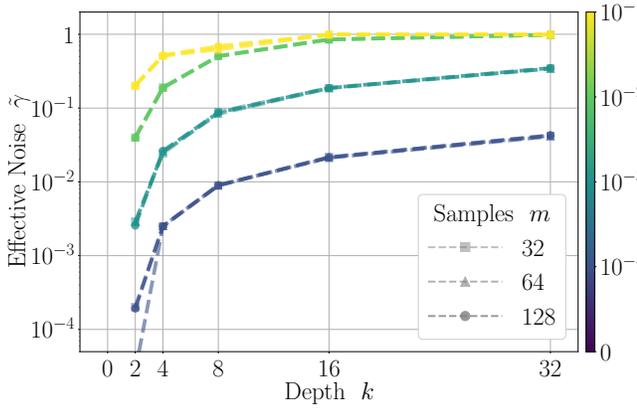}
		\subcaption{Effective noise $\tilde{\gamma}$ as a function of depth $k$.}
		\label{fig:plot_parameters_sic_povm_depth_noise}
	\end{subfigure}
	\begin{subfigure}[t]{0.49\columnwidth}
		\centering
		\includegraphics[width=1.01\textwidth]{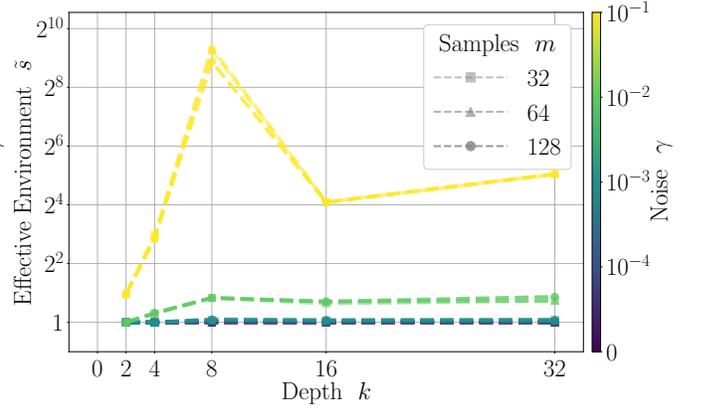}
		\subcaption{Effective environment $\tilde{s}$ as a function of depth $k$.}
		\label{fig:plot_parameters_sic_povm_depth_env}
	\end{subfigure}
	\captionsetup{justification=raggedright}
	\caption{Optimized parameters of effective noise $\tilde{\gamma}$ and effective environment dimension $\tilde{s}$ of an effective analytical model ${P}_{\mathcal{P}\gamma}^{(k)}(p|\tilde{\gamma},\tilde{s})$ for SIC-POVM measurement probability distributions, as a function of system parameters noise scale $\gamma$ and depth $k$, for system size $n=10$. Effective noise scales are shown to be smooth polynomial functions of noise scale, offset by increasing depth, and is consistent with noiseless behavior, $\lim_{\gamma \to 0}\tilde{\gamma} \to 0$. Effective environment dimensions remain constant at the minimal trivial $\tilde{s}=1$ environments, consistent with noiseless behavior, $\lim_{\gamma \to 0}\tilde{s} \to 1$, until sufficient noise causes a potentially exponential jump in environment dimension with noise. Negligible sample size effects are shown for different number of samples $m \in \{32,64,128\}$.}
	\label{fig:plot_parameters_sic_povm}
\end{figure}

Other behaviors to consider include the smoothness of the effective parameters, given they are expected to increase monotonically with noise scales and depth. The effective noise scale appears to have an intuitive form of a monotonic, polynomial function of noise scale. The effective environment dimension, being typically an integer, varies less smoothly with noise scale and depth. At low noise scales, the effective environment dimension remains trivially at one, indicating independent systems and environments in noiseless settings. At noise scales above a threshold, the effective environment dimension increases to scale exponentially with system size, indicative of highly entangled systems and environments in noisy settings. Further, the effective environment is largest at intermediate depths, potentially indicating where the noisy states are maximally Haar random mixed states.  In conclusion, although it is not confirmed that such trends are indicative of global optima of the effective model, the smoothly varying behaviors suggest that these effective parameters are physically valid, and representative of the simulated systems.

\end{document}